\documentclass[twocolumn,aps]{revtex4}
\usepackage{graphicx}
\usepackage{epsfig}
\usepackage{amsmath}

\renewcommand{\imath}{\mathrm{i}}
\newcommand{\figref}[1]{Fig.~\ref{#1}}
\newcommand{\figureref}[1]{Figure~\ref{#1}}

\newcommand{\eqnref}[1]{Eq.~\eqref{#1}}

\begin{document}

\title{Stabilizing continuous-wave output in semiconductor lasers by time-delayed feedback}
\author{Thomas Dahms}
\author{Philipp H{\"o}vel}
\author{Eckehard Sch{\"o}ll}\email{schoell@physik.tu-berlin.de}

\affiliation{Institut f{\"u}r Theoretische Physik, Technische
Universit{\"a}t Berlin, 10623 Berlin, Germany}
 
\date{\today}

\begin{abstract}
The stabilization of steady states is studied in a modified Lang-Kobayashi model of a semiconductor
laser. We show that multiple time-delayed feedback, realized by a Fabry-Perot resonator coupled to the laser, 
provides a valuable tool for the suppression of unwanted intensity pulsations, and leads to stable 
continuous-wave operation. 
The domains of control are calulated in dependence on the feedback strength, delay time (cavity round trip time), 
memory parameter (mirror reflectivity), latency time, feedback phase, and bandpass filtering,
Due to the optical feedback, multistable behavior can also occur in the form of delay-induced intensity pulsations
or other modes for certain choices of the control parameters. 
Control may then still be achieved by slowly ramping the injection current during turn-on.

\end{abstract}


\maketitle

\section{Introduction}
\label{sec:intro}
Time-delayed feedback methods have been widely used to control unstable dynamics in a variety of different fields 
\cite{SCH07}. In its original form, \textit{time-delay autosynchronisation} (TDAS) was introduced by Pyragas
\cite{PYR92}
to stabilize unstable periodic orbits embedded in a chaotic attractor. In TDAS, the control force is generated from
the difference of an output signal $s(t)$ and its counterpart $s(t-\tau)$ some time units $\tau$ ago. 
One groundbreaking advantage of TDAS is the noninvasiveness of the control, i.e., the control force vanishes on 
target orbit. This control scheme was extended by Socolar \textit{et al.}, by considering all integer multiples $m$ of the 
delay time $s(t-m\tau)- s(t-(m+1)\tau)$ weighted with a memory parameter $R^m$
(\textit{extended time-delay autosynchronisation}, ETDAS) \cite{SOC94}. 
This control scheme was invented in the context of optical systems like semiconductor lasers, where feedback can be 
realized all-optically, for instance, by a Fabry-Perot (FP) resonator, which naturally generates an ETDAS control force
by multiple reflections. Such all-optical noninvasive control has indeed been realized experimentally by coupling a 
multisection semiconductor laser with an external Fabry-Perot cavity to stabilize unstable steady states
\cite{SCH06a,WUE07}. 

A simple model describing a semiconductor laser with optical feedback from a single mirror was introduced by Lang and 
Kobayashi \cite{LAN80b}. The effect of time-delayed feedback (TDAS) on semiconductor lasers was investigated 
within Lang-Kobayashi (LK) type models \cite{SIM96,LIU97,TRO06,FLU07,FIE08}, as well as within more elaborate device
models  \cite{ROG99,SIM99a,FIS00a,BAU04,USH04,AHL06a,SCH06a} for various configurations, including Michelson
interferometers  providing a realization of TDAS. These findings were supported by experimental 
work \cite{ROG00, FIS00a, SCH06a}. Not only noninvasive control but also delay-induced instabilities
\cite{HEI01a,HEI03a} and high-dimensional chaos resulting from time-delayed feedback were studied \cite{FIS94}.
In particular, feedback-induced stationary external cavity modes and their bifurcations in a LK model for
a laser subject to resonant feedback from a Fabry-Perot resonator were treated within the TDAS approximation 
\cite{TRO06}.

\begin{figure}[ht]
\begin{center}
  \includegraphics[width=1.0\columnwidth]{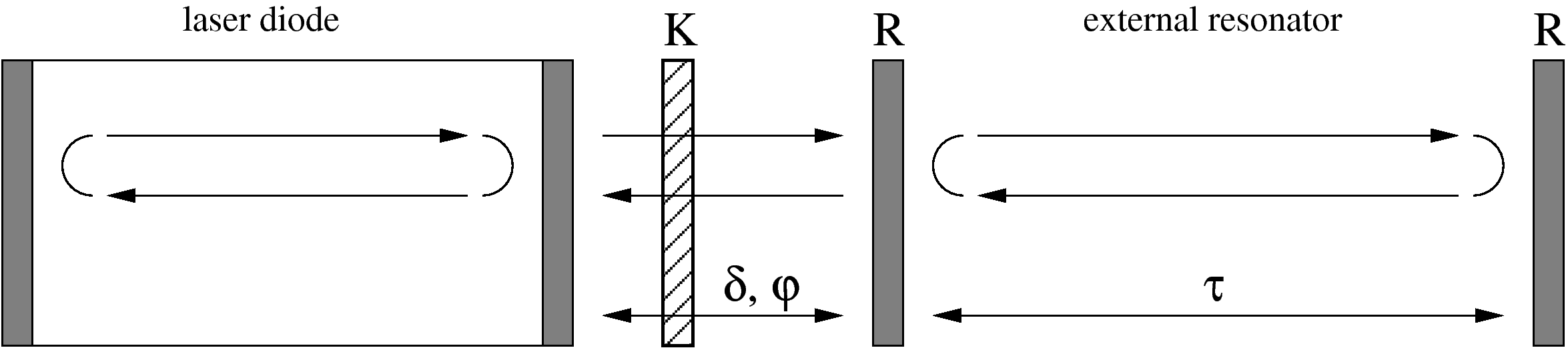}
  \end{center}
  \caption{Schematic diagram of a semiconductor laser with resonant feedback from a Fabry-Perot resonator. 
$K$ denotes an attenuator, $R$ is related to the mirror reflectivity of the external resonator, $\tau$ is 
the round trip time of the resonator, $\varphi$ and $\delta$ are phase shift and latency time due to the distance between laser and resonator.}
  \label{fig:laser}
\end{figure}

In this work, we will consider a modified LK model of a semiconductor laser coupled to a Fabry-Perot resonator
(see Fig.~\ref{fig:laser} for a schematic diagram of the system), and investigate how ETDAS can be successfully used to
stabilize unstable steady states, i.e., continuous wave (cw) emission, of the uncontrolled (uncoupled) system.  
It thus provides a systematic theoretical framework for the type of experimental configuration used in
Ref.~\cite{SCH06a}. We calculate the domains of control in dependence on various control parameters, latency, and
bandpass filtering, which extends our previous findings from a simple linear generic model on time-delayed feedback
control of a fixed point  \cite{HOE05,YAN06,DAH07}.
In addition, we present numerical simulations of the full nonlinear model and develop strategies for global control
of the fixed point even in case of delay-induced multistability. 
 
This paper is organized as follows: In Section~\ref{sec:model} we introduce a model of a semiconductor laser 
with internal passive dispersive feedback. This model allows for undamped relaxation oscillations for certain choices of
the parameters. Furthermore, optical feedback from a Fabry-Perot resonator is included.
In order to investigate the properties in the vicinity of the lasing fixed point, 
a linear stability analysis of the system is performed in Section~\ref{sec:linear}, both with and without 
external optical feedback. The numerical results are presented in Section~\ref{sec:results}, and the dependence upon
feedback strength, time delay, memory parameter, optical phase shift, and latency time is studied. In Section~\ref{sec:bandpass} a bandpass 
filter is applied to the feedback signal. Again, numerical results show the impact on stability of the lasing fixed
point. The full nonlinear dynamic behavior of the system is investigated in Section~\ref{sec:lk_multistability}. 
To overcome the problems of delay-induced multistability, simulations of a specifically designed
turn-on process allowing for successful control of the steady state
are discussed in Section~\ref{sec:turnon}. Finally, we finish with a conclusion in Section~\ref{sec:conclusion}.

\section{The model}
\label{sec:model}
Semiconductor lasers with external optical feedback from a mirror can be described by the 
Lang-Kobayashi model \cite{LAN80b}. In dimensionless form, it consists of two differential equations for the 
slowly varying amplitude (envelope) $E(t)$ of the complex electric field and the reduced carrier density (inversion) $n(t)$.

Here we consider a modification of the LK equations appropriate for multisection semiconductor lasers with
an internal passive dispersive reflector~\cite{TRO00}. This is modeled by a gain function $k(n)$ 
depending upon the internal dispersive feedback from the Bragg grating. 
Such a laser structure allows for more complex dynamic behavior including self-sustained 
relaxation oscillations (intensity pulsations) generated by Hopf bifurcations, as has been shown 
in the framework of traveling wave laser models \cite{BAU04,SCH06a}. We are interested in the regime 
above a supercritical
Hopf bifurcation where the fixed point in the uncontrolled system is unstable. Combining the rate equation for the carrier
density from Ref.~\cite{TRO00} with the rate equation for the complex electric field, we obtain the following form
of modified LK equations:
\begin{subequations}
\label{eqn:lk_tronciu}
\begin{eqnarray}
  \frac{d E}{d t} & = & \frac{T}{2} \left( 1+ \imath \alpha \right) n E - E_{b}(t) ,\\
  \frac{d n}{d t} & = & I -n- (1+n) k(n) \left| E \right|^{2},
\end{eqnarray}
\end{subequations}
where $\alpha$ denotes the linewidth enhancement factor, $I$ is the reduced excess injection current, 
$T$ is the time scale ratio of the carrier lifetime $\tau_c$ and the photon 
lifetime $\tau_p$, and $E_{b}(t)$ denotes the feedback term, which will be described in detail later. 

\begin{figure}[ht]
\begin{center}
  \includegraphics[width=1.0\columnwidth]{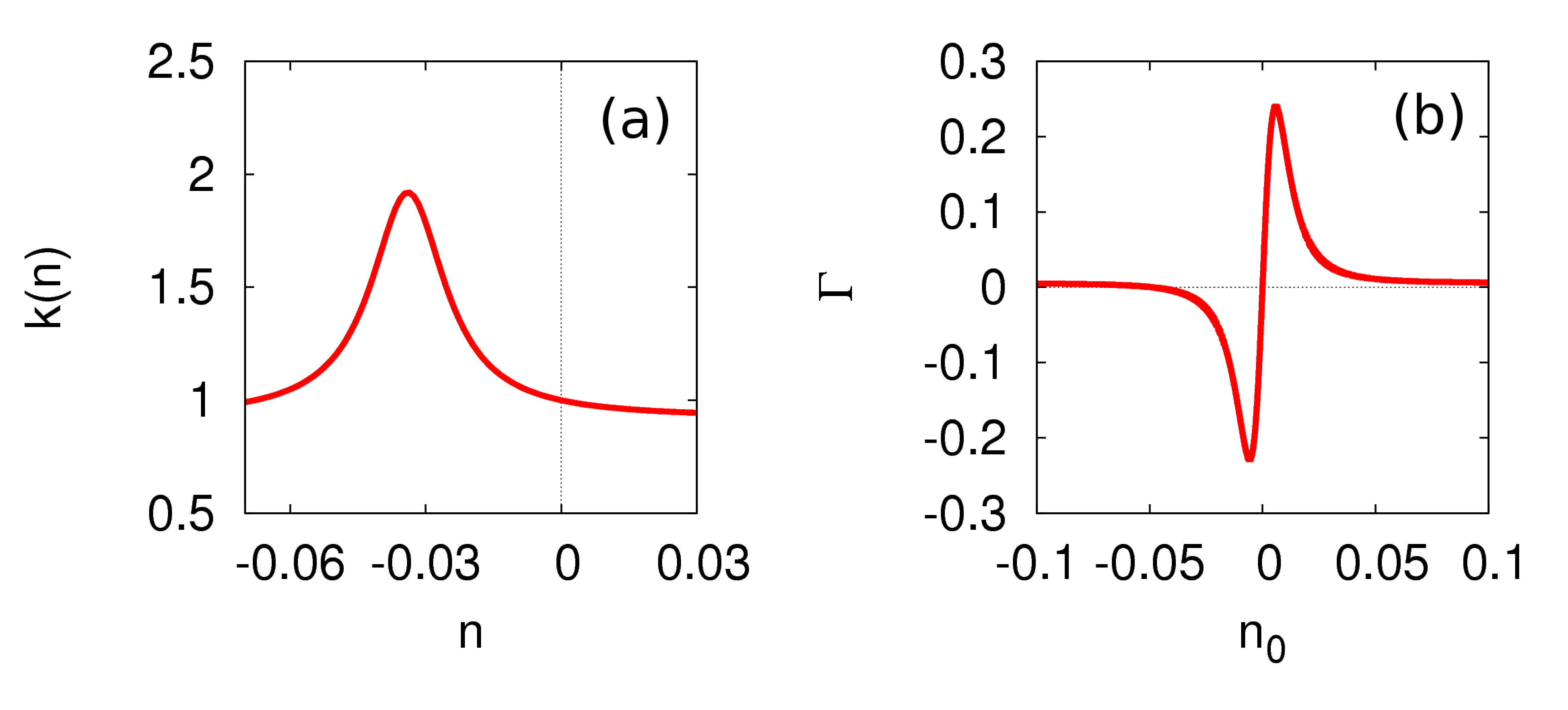}
  \end{center}
  \caption{(a) Typical shape of the gain function $k(n)$ used in the modified LK model \eqnref{eqn:lk_tronciu}.
Parameters: $A=1$, $W=0.02$, and $n_{0}=-0.034$. (b) Dependence of the damping rate $\Gamma$ on the choice of the
detuning parameter $n_{0}$ in the function $k(n)$ according to Eqs.~(\ref{eqn:lk_kvonn}) and (\ref{eqn:lk_Gamma}).}
  \label{fig:lk_kvonn}
\end{figure}

Here we have scaled (i) time $t$ by the carrier lifetime $\tau_c$, (ii) carrier density $n$
(in excess of the threshold carrier density) by the inverse of the 
differential gain $G_N$ times $\tau_p$, and (iii) electric field $E$ by
$(\tau_c G_N)^{-1/2}$, so that all variables are dimensionless. Moreover, the variables $I$ and $n$ 
are linearly transformed such that both are zero at the laser threshold. Note that here, as in Ref. 
~\cite{TRO00}, time is scaled by $\tau_c$ rather than by $\tau_p$, as often used elsewhere (see, e.g. \cite{ALS96}).
The function $k(n)$, which models 
the internal dispersive feedback, is chosen as a Lorentzian, as proposed in Ref.~\cite{TRO00}:
\begin{equation}
k(n)=k_{0} + \frac{A W^{2}}{4\left(n-n_{0}\right)^{2}+W^{2}} ,
\label{eqn:lk_kvonn}
\end{equation}
where $A$ denotes the height, $W$ is the width, and $n_{0}$ is the position of the resonance. 
The parameter $k_{0}$ is chosen such that $k(0)=1$ at the laser threshold. A typical curve $k(n)$ is shown 
in \figref{fig:lk_kvonn}(a). Througout the following we will use the parameters $A=1$, $W=0.02$, and $n_0 = -0.034$.

In Eqs.~(\ref{eqn:lk_tronciu}), the feedback term $E_{b}(t)$ has not been specified yet. In the following, 
we introduce the feedback term such that it models a Fabry-Perot resonator. As opposed to the original 
LK model where only a single external mirror is considered, we take an external FP resonator with multiple reflections
into account: 
\begin{eqnarray}
 \lefteqn{E_{b}(t)} \label{eqn:lk_fp_feedback}\\
 & = & K e^{-i\varphi} \sum_{m=0}^{\infty} R^{m} \left[ E(t-\delta-m \tau) - E(t-\delta-(m+1)\tau) \right] \nonumber \\
 & = & K e^{-i\varphi} \left[ E(t-\delta) - E(t-\delta-\tau) \right] + R E_{b}(t-\tau) \nonumber .
\end{eqnarray}
$K$ is the feedback strength, $\tau$ is the delay time (cavity round trip time), $R$ is a memory parameter 
(mirror reflectivity), $\delta$ denotes the latency time originating from a single round trip between the laser and the
resonator (see Fig.~\ref{fig:laser}), and $\varphi$ is the feedback phase which results from the associated optical
phase shift.

The latency time $\delta$, i.e., the propagation time between the laser and the FP, 
is correlated to the phase $\varphi$ by the relation $\varphi=\Omega_{0}\delta$, 
where $\Omega_{0}$ is the frequency of the emitted light.
However, we consider the two parameters $\varphi$ and $\delta$ 
as independent variables because the phase $\varphi$ can be tuned by subwavelength changes of the separation
between laser and FP, on which scale the slowly varying amplitude $E$, which depends upon $\delta$, does not change.
The effect of latency in time-delayed feedback was already studied in a general context in
Refs.~\cite{JUS99b,HOE03,HOE05,DAH07}.

Throughout this work we use resonant feedback from the FP, i.e., the additional phase shift of the electric field  
by round trips in the FP resonator, denoted by $\phi$, is assumed to be $2\pi m$, where $m$ is an integer. Otherwise
the control term would be modified by an additional factor $e^{-\imath\phi}$ in the second term, 
and would not vanish for $E(t-\delta) = E(t-\delta-\tau)$.

\section{Linearization around the lasing fixed point}
\label{sec:linear}
To investigate the stability of the cw laser emission, we will first consider the system without the feedback term.
The uncontrolled system given by Eqs.~\eqref{eqn:lk_tronciu} has a trivial fixed point at $(n=I,E=0)$. This fixed point is 
not of interest, because it describes a non-lasing state, i.e., the electric field is zero.

Another fixed point, which describes a lasing state is located at $(n=0,E=\sqrt{I} e^{\imath \psi})$. Now, the phase of 
the electric field can be arbitrarily fixed to $\psi=0$, such that $e^{\imath \psi}=1$ (solitary laser mode). This 
leads to symmetry breaking of the rotational (S1) symmetry of Eqs.~\eqref{eqn:lk_tronciu} with respect to complex E.

Using the abbreviations
\begin{subequations}
\begin{eqnarray} 
 E(t) & = & \sqrt{T} \left[ \Omega_{0} + x(t) + \imath y(t) \right] \label{eqn:lk_E_abkuerzung} ,\\
 \Omega_{0} & = & \sqrt{\frac{I}{T}} ,\\
 \Gamma & = & \frac{1}{T} \left[ 1 + I \left( 1 + \left.\frac{d k}{d n}\right|_{n=0}\right) \right] \label{eqn:lk_Gamma},
\end{eqnarray}
\end{subequations}
with real-valued $x$ and $y$, the fixed point is located at $(n=0,x=0,y=0)$. A linearization around this point leads to the following system of equations:
\begin{equation}
\left( \begin{array}{c}
  \dot{n}(t)\\
  \dot{x}(t)\\
  \dot{y}(t)
\end{array} \right) =
\left( \begin{array}{ccc}
  -2\Gamma & -4\Omega_{0} & 0 \\
  \Omega_{0} & 0 & 0 \\
  \Omega_{0}\alpha & 0 & 0
\end{array} \right)
\left( \begin{array}{c}
  n(t)\\
  x(t)\\
  y(t)
\end{array} \right) .
\label{eqn:lk_linear}
\end{equation}
Note that an additional time scale transformation $t \rightarrow (T/2)t$ was performed to eliminate the parameter $T$
from the equations. Thus the rescaled time $t$ is related to the physical time $s$ by
\begin{eqnarray}\label{eq:scale} 
 t = \frac{s}{\tau_c}\;\frac{T}{2} = \frac{s}{2\tau_p}.
\end{eqnarray}
Also, the function $k(n)$ is no longer present in the linearized equations. Instead, a parameter $\Gamma$ was introduced
that includes the differential gain $\mathrm{d}k/\mathrm{d}n$ evaluated at the fixed point. The relation between
$\Gamma$ and the parameter $n_{0}$ used in $k(n)$  according to Eqs.~(\ref{eqn:lk_kvonn}) and (\ref{eqn:lk_Gamma}) is
shown in \figref{fig:lk_kvonn}(b). The stability properties of the lasing fixed point are evaluated within this
linearized model in the following. 

In order to investigate the stability of the fixed point in the uncontrolled system, we consider the characteristic
equation for the eigenvalue $\Lambda$ which is given by:
\begin{subequations}
\begin{eqnarray}
 0 & = & \det{ \left[ \left( \begin{array}{ccc}
  -2\Gamma & -4\Omega_{0} & 0 \\
  \Omega_{0} & 0 & 0 \\
  \Omega_{0}\alpha & 0 & 0
\end{array} \right) - \Lambda \mathbf{Id} \right] } \\
 & = & \Lambda\left( \Lambda^{2} + 2 \Gamma \Lambda + 4 \Omega_{0}^{2} \right),
\end{eqnarray}
\end{subequations}
where $\mathbf{Id}$ is the identity matrix. The solutions of this equation are given by
\begin{subequations}
\begin{eqnarray}
 \Lambda &=& 0, \\
 \Lambda &=& -\Gamma \pm \imath \sqrt{4\Omega_{0}^{2}-\Gamma^{2}} . \label{eqn:lk_eigenvalues_uncontrolled}
\end{eqnarray}
\end{subequations}
The Goldstone mode $\Lambda=0$ does not contain information about the stability of the fixed point and is only present due to the rotation symmetry. Therefore, only the second solution, given by \eqnref{eqn:lk_eigenvalues_uncontrolled}, is of interest.

Note that the imaginary part $\sqrt{4\Omega_{0}^{2}-\Gamma^{2}}$, which corresponds to the frequency of the relaxation oscillation, exists only for $|2\Omega_{0}| > |\Gamma|$. Under this condition the fixed point is a focus. Since $-\Gamma$ is the real part of the eigenvalue $\Lambda$, the focus is stable for any $\Gamma > 0$ and unstable otherwise. Starting with an unstable focus ($\Gamma < 0$), we investigate the effects of the extended time-delayed feedback control on the stability of the fixed point in the following.

Adding the control term Eq.(\ref{eqn:lk_fp_feedback}) with properly rescaled $K$ to the linearized equations leads to
the following system:
\begin{widetext}
\begin{eqnarray}
\left( \begin{array}{c}
  \dot{n}(t)\\
  \dot{x}(t)\\
  \dot{y}(t)
\end{array} \right) & = &
\left( \begin{array}{ccc}
  -2\Gamma & -4\Omega_{0} & 0 \\
  \Omega_{0} & 0 & 0 \\
  -\Omega_{0}\alpha & 0 & 0
\end{array} \right)
\left( \begin{array}{c}
  n(t)\\
  x(t)\\
  y(t)
\end{array} \right) - K \left( \begin{array}{ccc}
  0 & 0 & 0 \\
  0 & \cos{\varphi} & \sin{\varphi} \\
  0 & -\sin{\varphi} & \cos{\varphi}
\end{array} \right) \nonumber \\
& & \times
\left( \begin{array}{c}
  0\\
  \sum_{m=0}^{\infty} R^{m} \left[ x(t-\delta-m \tau) - x(t-\delta-(m+1)\tau) \right]\\
  \sum_{m=0}^{\infty} R^{m} \left[ y(t-\delta-m \tau) - y(t-\delta-(m+1)\tau) \right]
\end{array} \right) .
\end{eqnarray}
Using an exponential ansatz $\exp{(\Lambda t)}$ for all three variables $x$, $y$, and $n$
leads to the characteristic equation 
\begin{eqnarray}
 0 & = & \left( 2 \Gamma + \Lambda \right) \left[ \left( K e^{-\Lambda\delta}\frac{1-e^{-\Lambda \tau}}{1- R e^{-\Lambda \tau}} \right)^{2} + \Lambda^{2} + 2 \Lambda K e^{-\Lambda\delta}\frac{1-e^{-\Lambda \tau}}{1- R e^{-\Lambda \tau}} \cos{\varphi }\right] \nonumber \\
 & & + 4 \Omega_{0}^{2} \left( \Lambda + K e^{-\Lambda\delta}\frac{1-e^{-\Lambda \tau}}{1- R e^{-\Lambda \tau}} \cos{\varphi} + \alpha K e^{-\Lambda\delta}\frac{1-e^{-\Lambda \tau}}{1- R e^{-\Lambda \tau}} \sin{\varphi} \right).
 \label{eqn:EIGENVALUES_LK}
\end{eqnarray}
\end{widetext}
In our simulations, we use the following parameters, which were chosen close to the values of Ref.~\cite{TRO00}: $\Omega_{0} =0.06$, $\alpha =5$, $\Gamma =-0.01$ (corresponding to $T=500$,  $I = 1.8$,  $A = 1$, $W=0.02$, $n_{0} = -0.033933$, and $k_{0} = 0.993075$). Thus the intrinsic period of the uncontrolled unstable focus is $T_{0} \approx \pi/\Omega_{0} \approx 52$.

Note that the value of $k_{0}$ is determined by the other three parameters via the constraint $k(0)=1$. 
With the values given here, the domain of control will be investigated in the $(K,\tau)$- and $(K,R)$-plane in the following.

In the special case of $\varphi=0$ and $\delta=0$ the boundary of the control domain can be obtained analytically in a similar way as in Ref.~\cite{DAH07} for the generic normal form model. The characteristic equation \eqref{eqn:EIGENVALUES_LK} can be factorized into
\begin{subequations}
\begin{eqnarray}
0 &=& \Lambda + K \frac{1-e^{-\Lambda \tau}}{1-R e^{-\Lambda \tau}} , \label{eqn:analytisch1} \\
0 &=& \left( 2 \Gamma + \Lambda \right) \left( \Lambda + K \frac{1-e^{-\Lambda \tau}}{1-R e^{-\Lambda \tau}}\right) + 4 \Omega_{0}^{2} , \label{eqn:analytisch2}
\end{eqnarray}
\end{subequations}
where Eq.~\eqref{eqn:analytisch1} corresponds to the Goldstone mode in the uncontrolled system and therefore only Eq.~\eqref{eqn:analytisch2} is considered for the stability analysis. Separating Eq.~\eqref{eqn:analytisch2} into real and imaginary part and using the Hopf condition $\mathrm{Re}(\Lambda)=0$ yields after some trigonometric manipulations the parametric representation of the control domain boundaries in the $(K,\tau)$ parameter plane, parametrized by the imaginary part $\mathrm{Im}(\Lambda)=:q$ of the complex eigenvalue:
\begin{subequations}
\begin{eqnarray}
K(q) &=&  \frac{\left(1+R\right)\left[q^{4}+16\Omega_{0}^{4}+4q^{2}\left(\Gamma^{2}-2\Omega_{0}^{2}\right)\right]}{-16\Gamma\Omega_{0}^{2}},
\label{eqn:lk_k_tau_k} \\
\tau_{1}(q) &=& \frac{\arcsin{\left[a(q)\right]} + 2 m \pi}{q} ,\label{eqn:lk_k_tau_tau}\\
\tau_{2}(q) &=& \frac{-\arcsin{\left[a(q)\right]} + \left(2m+1\right) \pi}{q} ,
\end{eqnarray}
\end{subequations}
with
\begin{eqnarray}
 a(q)&=& \left\{ K(q) q \left(R-1\right) \right.  \nonumber \\
& & \ \left. \left[q^{2}+4\left(\Gamma-\Omega_{0}\right)\left(\Gamma+\Omega_{0}\right)\right]\right\} / \nonumber \\
& & \left\{\left(K(q)^{2}+R^{2} q^{2}\right)\left(q^{2}+4\Gamma^{2}\right) \right. \nonumber \\
& & \ \left. - 8\Omega_{0}^{2} R \left(R q^{2} - 2 K(q) \Gamma\right) +16 R^{2} \Omega_{0}^{4}\right\}, \nonumber
\end{eqnarray}
where $m$ is any nonnegative integer. This parameter had to be introduced due to the multiple leaves of the $\arcsin$
function.

\section{Numerical results}
\label{sec:results}
The stability of the lasing fixed point is given by Eq. \eqref{eqn:EIGENVALUES_LK}.
We solved this equation via Newton's method. Since the transcendental
equation has an infinite number of roots, we scanned the complex plane as
initial conditions of the root-finding algorithm to locate the eigenvalue
with the largest real part.
The parameter space consisting of $K$, $\tau$, $R$, $\varphi$, and $\delta$ is five-dimensional for fixed $\Gamma$ and $\Omega_{0}$. To visualize the domain of control, we consider two-dimensional sections of this parameter space.

\begin{figure}[ht]
	 \includegraphics[width=1.0\columnwidth]{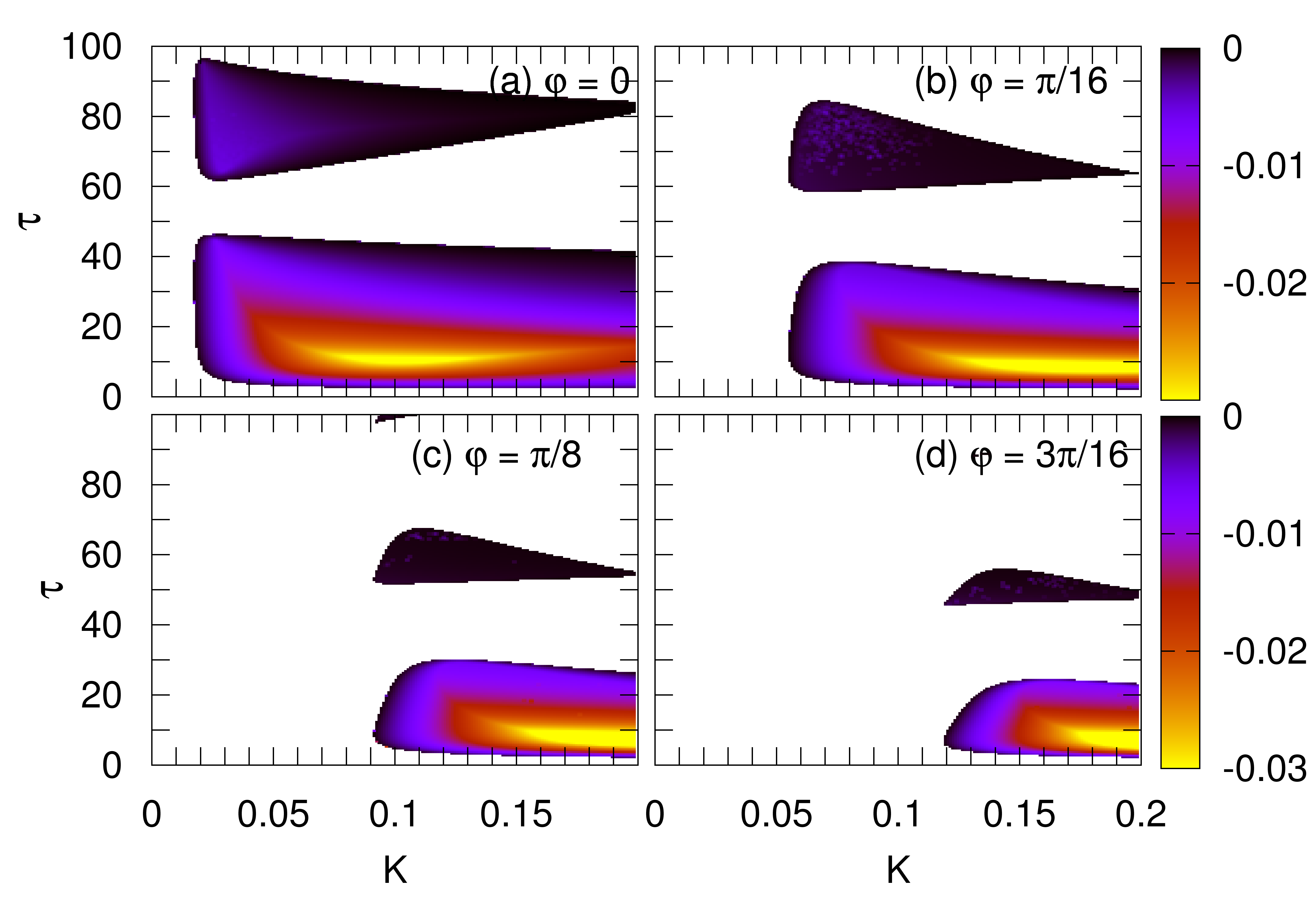}
	\caption{(Color online) Domain of control according to Eq.~\eqref{eqn:EIGENVALUES_LK} in the $(K,\tau)$-plane
for different values of $\varphi$. The greyscale (color code) denotes the largest real part $\mathrm{Re}(\Lambda)$ of
the eigenvalues $\Lambda$, only negative values are plotted. Panels (a), (b), (c), and (d) correspond to $\varphi=0$,
$\pi/16$, $\pi/8$, and $3\pi/16$, respectively. Other parameters: $\Gamma=-0.01$, $\Omega_{0}=0.06$, $\alpha=5$,
$R=0.7$, and $\delta=0$.}
	\label{fig:lk_k_tau_domain_phi}
\end{figure}
In \figref{fig:lk_k_tau_domain_phi}, the domain of control is shown in the $(K,\tau)$-plane for different values of the
phase: $\varphi=0$, $\pi/16$, $\pi/8$, and $3\pi/16$ in panels (a), (b), (c), and (d), respectively. The greyscale
(color code) denotes the largest real part of the eigenvalues and is therefore a measure of stability. Note that only
values of $\mathrm{Re}(\Lambda)<0$ are plotted, thus the shaded regions correspond to a stable lasing fixed point, i.e.,
a stable cw output. The control domains form tongues separated by (white) regions of no control around $\tau=n T_{0}$
with $n$ integer, just as in the generic model studied in Refs.~\cite{HOE05,DAH07}. It can be seen that the domain of
control shrinks with increasing phase.
Here, the domains of control are cut off by boundaries from the upper left and right for increasing phase, leading to
overall smaller regions of stability. The tongues of stabilization are also slightly distorted towards smaller values of
$\tau$. Additionally, in this picture, the regions of optimum stability, denoted by bright (yellow) color, are shifted
towards larger values of the feedback gain $K$.

\begin{figure}[ht]
	\includegraphics[width=1.0\columnwidth]{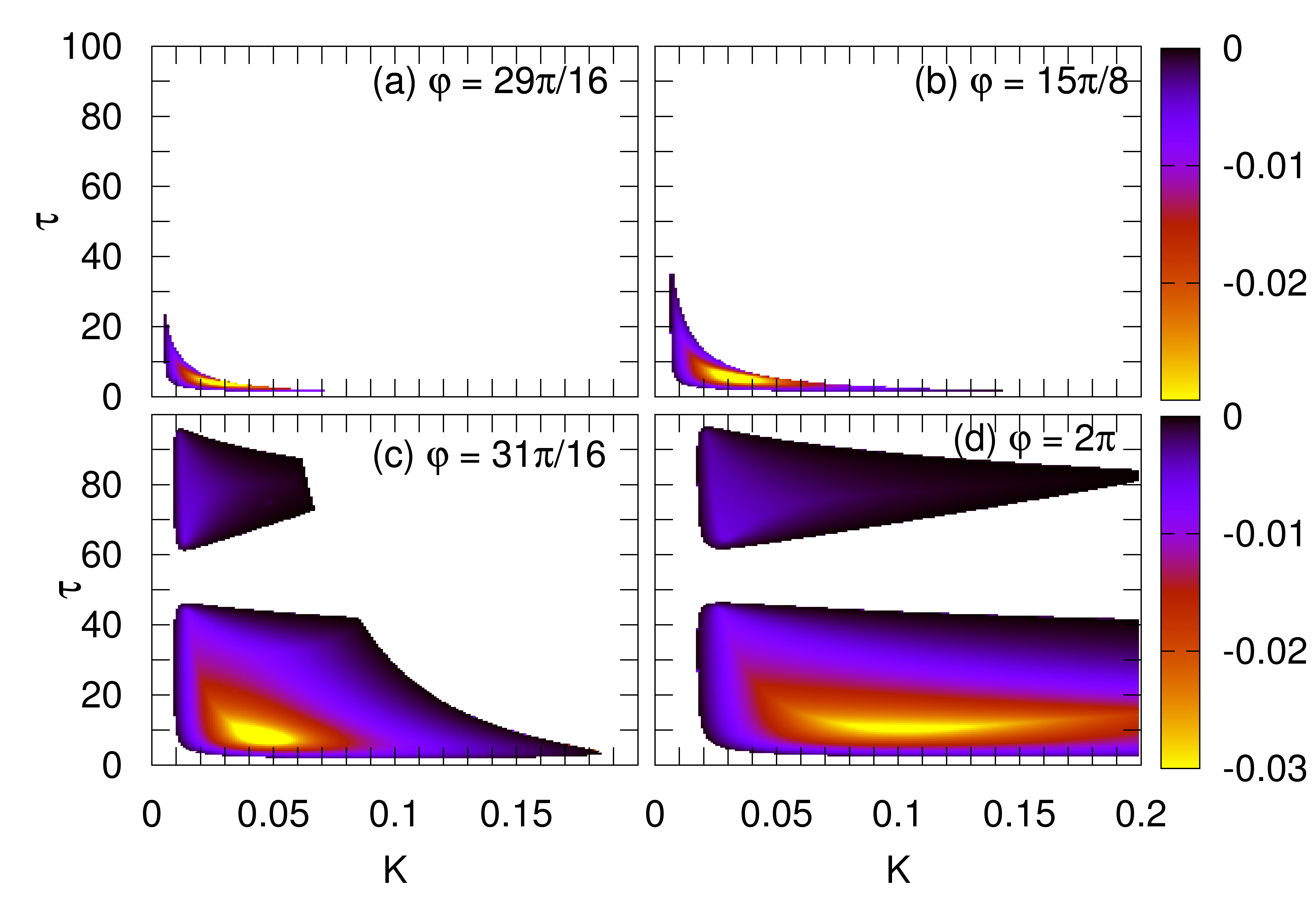}
	\caption{(Color online) Domain of control in the $(K,\tau)$-plane for different values of $\varphi$. The
greyscale (color code) denotes the largest real part $\mathrm{Re}(\Lambda)$ of the eigenvalues $\Lambda$, only negative
values are plotted. Panels (a), (b), (c), and (d) correspond to $\varphi=29\pi/16$, $15\pi/8$, $31\pi/16$, and $2\pi$,
respectively. Other parameters as in Fig.~\ref{fig:lk_k_tau_domain_phi}.}
	\label{fig:lk_k_tau_domain_phi2}
\end{figure}
\figureref{fig:lk_k_tau_domain_phi2} shows the domain of control in the $(K,\tau)$-plane for values of $\varphi$ in the
range $[1.8\pi,2\pi]$, which corresponds to negative phases $[-0.2\pi,0]$ because the feedback of
Eq.~(\ref{eqn:lk_fp_feedback}) shows a $2\pi$-periodicity in the phase variable.
Note that the minimum feedback is only slightly changed by decrease of the phase. However, at larger values of $K$ the
control domain is cut off by an asymptotic boundary from the upper right, leading to a strong decrease of the maximum
feedback gain with decreasing $\varphi$. In addition, the successive tongues of stabilization at larger values of $\tau$
vanish altogether. Also, the region of optimum stability, denoted by bright (yellow) color shrinks for decreasing
$\varphi$.

\begin{figure}[ht]
	\includegraphics[width=1.0\columnwidth]{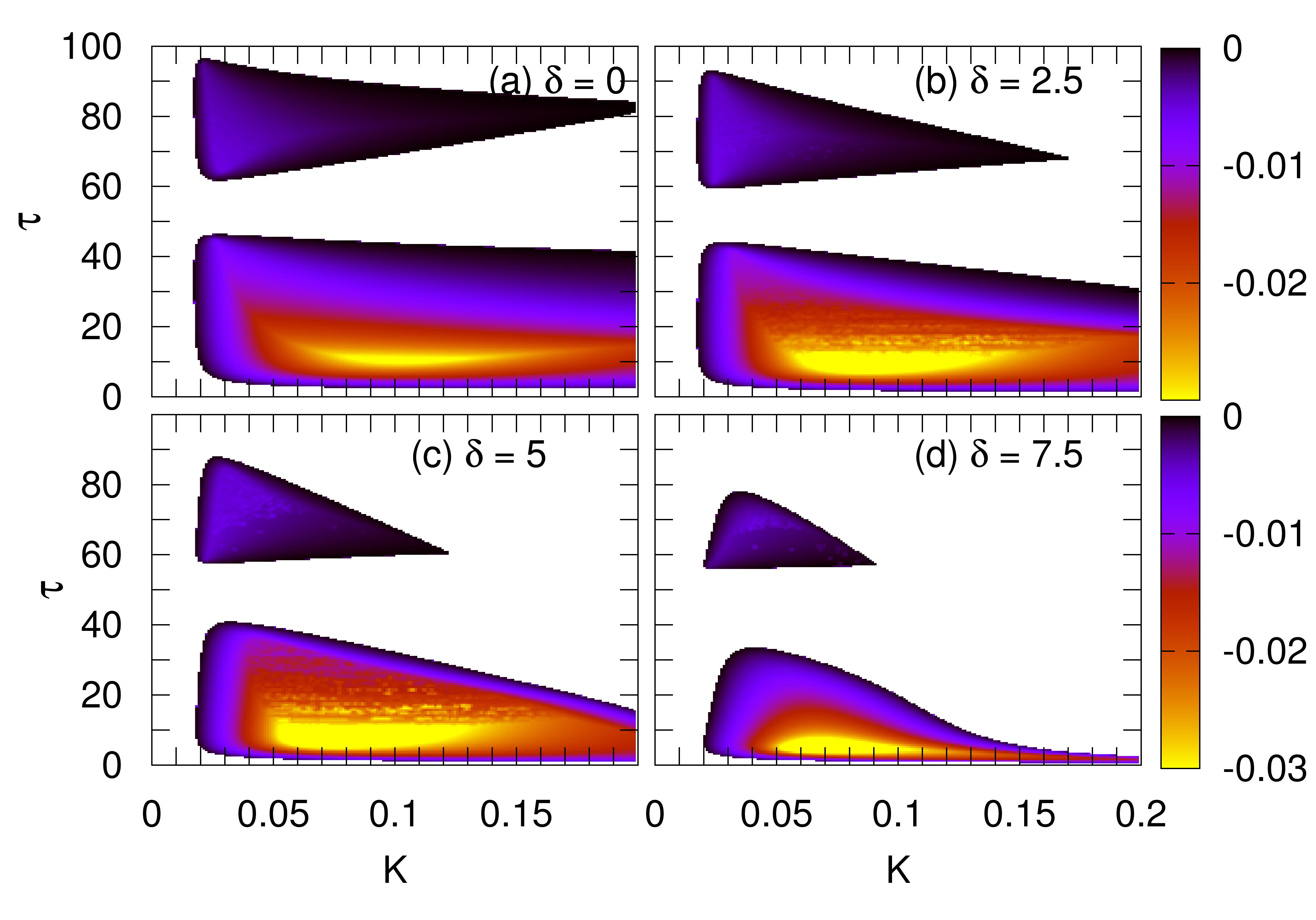}
	\caption{(Color online) Domain of control in the $(K,\tau)$-plane for different values of $\delta$ and fixed
$\varphi=0$. The greyscale (color code) denotes the largest real part $\mathrm{Re}(\Lambda)$ of the eigenvalues
$\Lambda$, only negative values are plotted. Panels (a), (b), (c), and (d) correspond to $\delta=0$, $2.5$, $5$, and
$7.5$, respectively. Other parameters as in Fig.~\ref{fig:lk_k_tau_domain_phi}.}
	\label{fig:lk_k_tau_domain_latency}
\end{figure}
Next, we will investigate the role of the latency time in the $(K,\tau)$-plane. In \figref{fig:lk_k_tau_domain_latency},
the domain of control in the $(K,\tau)$-plane is depicted for different values of the latency time, i.e., $\delta=0$,
$2.5$, $5$, and $7.5$, and fixed $\varphi=0$. For larger latency times, the domains of control shrink and it can also be
observed that they are bent down towards smaller values of the time delay $\tau$. 
Note that the regions of optimum stability, denoted by bright (yellow) color, are only slightly affected by the change
of the latency time.

\begin{figure}[ht]
	\includegraphics[width=1.0\columnwidth]{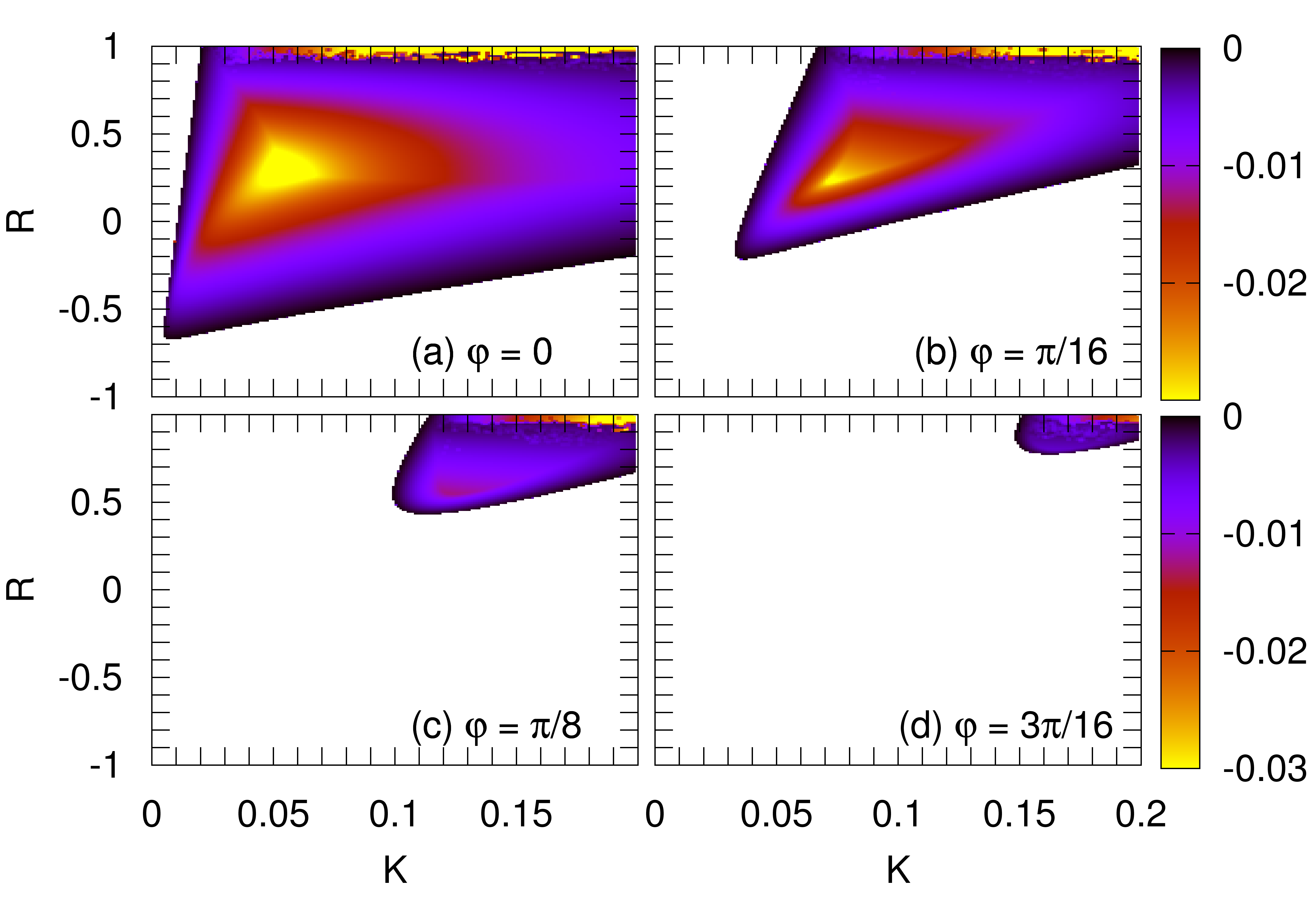}
\caption{(Color online) Domain of control in the $(K,R)$-plane for different values of $\varphi$ and fixed
optimum time delay $\tau=26$. The greyscale (color code) denotes the largest real part $\mathrm{Re}(\Lambda)$ of the
eigenvalues $\Lambda$, only negative values are plotted. Panels (a), (b), (c), and (d) correspond to $\varphi=0$,
$\pi/16$, $\pi/8$, and $3\pi/16$, respectively.  Other parameters as in Fig.~\ref{fig:lk_k_tau_domain_phi}.}
	\label{fig:lk_k_r_domain_phi}
\end{figure}

All figures shown here were obtained for a fixed value of the memory parameter $R=0.7$ as used in Ref.~\cite{SCH06a}. To
investigate the dependence of the control on $R$, we display the domains of control in the $(K,R)$-plane for different
values of the phase ($\varphi=0$, $\pi/8$, $\pi/4$, and $3\pi/8$) and fixed time delay $\tau=26$ in
\figref{fig:lk_k_r_domain_phi}. This value of $\tau$ was chosen based on the results from the generic model considered
in Refs.~\cite{HOE05,DAH07}, where it was shown that the optimum time delay is given by
$\tau=T_{0}/2=\pi/\mathrm{Im}(\Lambda_0)$, where $\Lambda_0$ denotes the eigenvalue of the uncontrolled system. In the
LK model, the imaginary part of the eigenvalues in the uncontrolled system is given by
\eqnref{eqn:lk_eigenvalues_uncontrolled}, i.e., $\mathrm{Im}(\Lambda_0) =  \sqrt{4\Omega_{0}^{2}-\Gamma^{2}}$. This
leads to an optimum time delay:
\begin{equation}
 \tau_{opt} = \frac{\pi}{\sqrt{4\Omega_{0}^{2}-\Gamma^{2}}} ,
 \label{eqn:tau_opt}
\end{equation}
which yields for our parameters $\tau_{opt} \approx 26$.

Now, in \figref{fig:lk_k_r_domain_phi}, it can be seen that the domain of control in the $(K,R)$-plane has maximum size
for $\varphi=0$ for this choice of the time delay $\tau$. [See panel \figref{fig:lk_k_r_domain_phi}(a).] For increasing
phase, the domain of control shrinks while moving to the upper right. Stability is then only achieved in a small region
at large values of $K$ and $R$. [See panels \figref{fig:lk_k_r_domain_phi}(b) to (d).]

\begin{figure}[ht]
	\includegraphics[width=1.0\columnwidth]{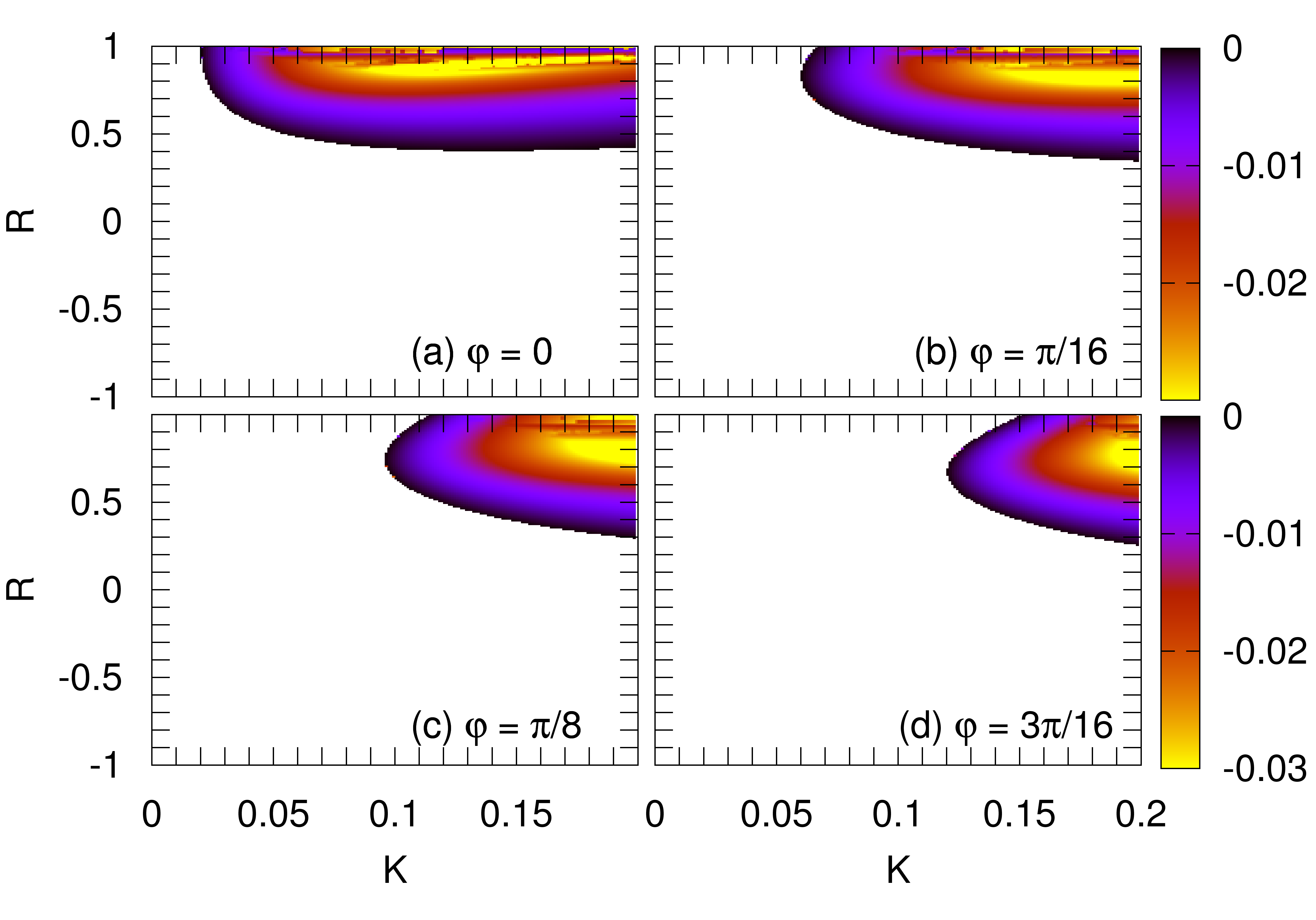}
	\caption{(Color online) Domain of control in the $(K,R)$-plane for different values of $\varphi$ and fixed time
delay $\tau=5$. The greyscale (color code) denotes the largest real part $\mathrm{Re}(\Lambda)$ of the eigenvalues
$\Lambda$, only negative values are plotted. Panels (a), (b), (c), and (d) correspond to $\varphi=0$, $\pi/16$, $\pi/8$,
and $3\pi/16$, respectively. Other parameters as in Fig.~\ref{fig:lk_k_tau_domain_phi}.}
	\label{fig:lk_k_r_domain_phi2}
\end{figure}
For the generic model as used in Ref.~\cite{HOE05}, it was shown that stability is enhanced in the case of nonzero phase
if the delay time is chosen different from its optimum value of half the intrinsic period $T_{0}$. Therefore, we
consider the domain of control in the $(K,R)$-plane for a value of $\tau=5 \approx 0.1 T_{0}$ in
\figref{fig:lk_k_r_domain_phi2} and for the same values of the phase $\varphi$ as in \figref{fig:lk_k_r_domain_phi}.
Here, the domain of control is already smaller in the case $\varphi=0$.
Increasing the phase, the domain of control shrinks in $K$-direction, but is enhanced slightly in the $R$-direction,
leading to an overall larger domain of control than for the optimum $\tau=26$. However, in the generic normal form model
\cite{DAH07}, the benefit of the non-optimum time delay was much more pronounced.

\begin{figure}[ht]
	\includegraphics[width=1.0\columnwidth]{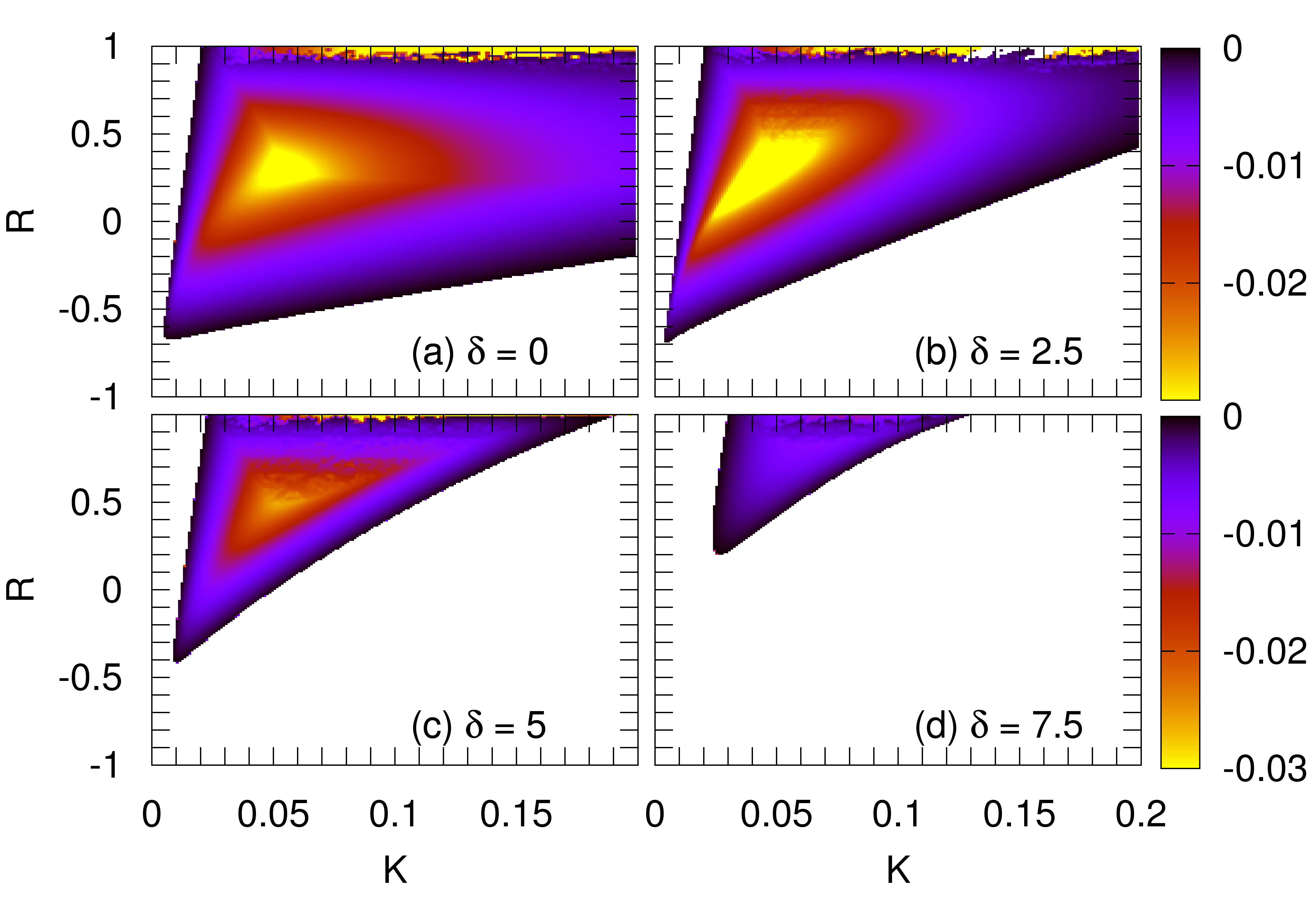}
	\caption{(Color online) Domain of control in the $(K,R)$-plane for different values of $\delta$. The time delay
and feedback phase are fixed at optimum $\tau=26$ and $\varphi=0$, respectively. The greyscale (color code) denotes the
largest real part $\mathrm{Re}(\Lambda)$ of the eigenvalues $\Lambda$, only negative values are plotted. Panels (a),
(b), (c), and (d) correspond to $\delta=0$, $2.5$, $5$, and $7.5$, respectively.  Other parameters as in
Fig.~\ref{fig:lk_k_tau_domain_phi}.}
	\label{fig:lk_k_r_domain_delta}
\end{figure}
Similar investigations as performed here for nonzero phase, are reported in the following for nonzero latency time
$\delta$ and fixed $\varphi=0$. In \figref{fig:lk_k_r_domain_delta}, the domain of control in the $(K,R)$-plane is shown
for different values of the latency time and fixed $\tau=26$. For better comparison, we choose the same latency times as
in \figref{fig:lk_k_tau_domain_latency}. It can be observed that the domain of control shrinks for increasing $\delta$.
The lower right boundary is shifted upwards. Therefore, control is only possible for large values of $R$ if $\delta$ is
large. The minimum feedback gain is also increased, since the left boundary is shifted to the right. The regions of
optimum stability, denoted by bright (yellow) color shrink for increasing $\delta$, leading to a deterioration of the
control.

\begin{figure}[ht]
	\includegraphics[width=1.0\columnwidth]{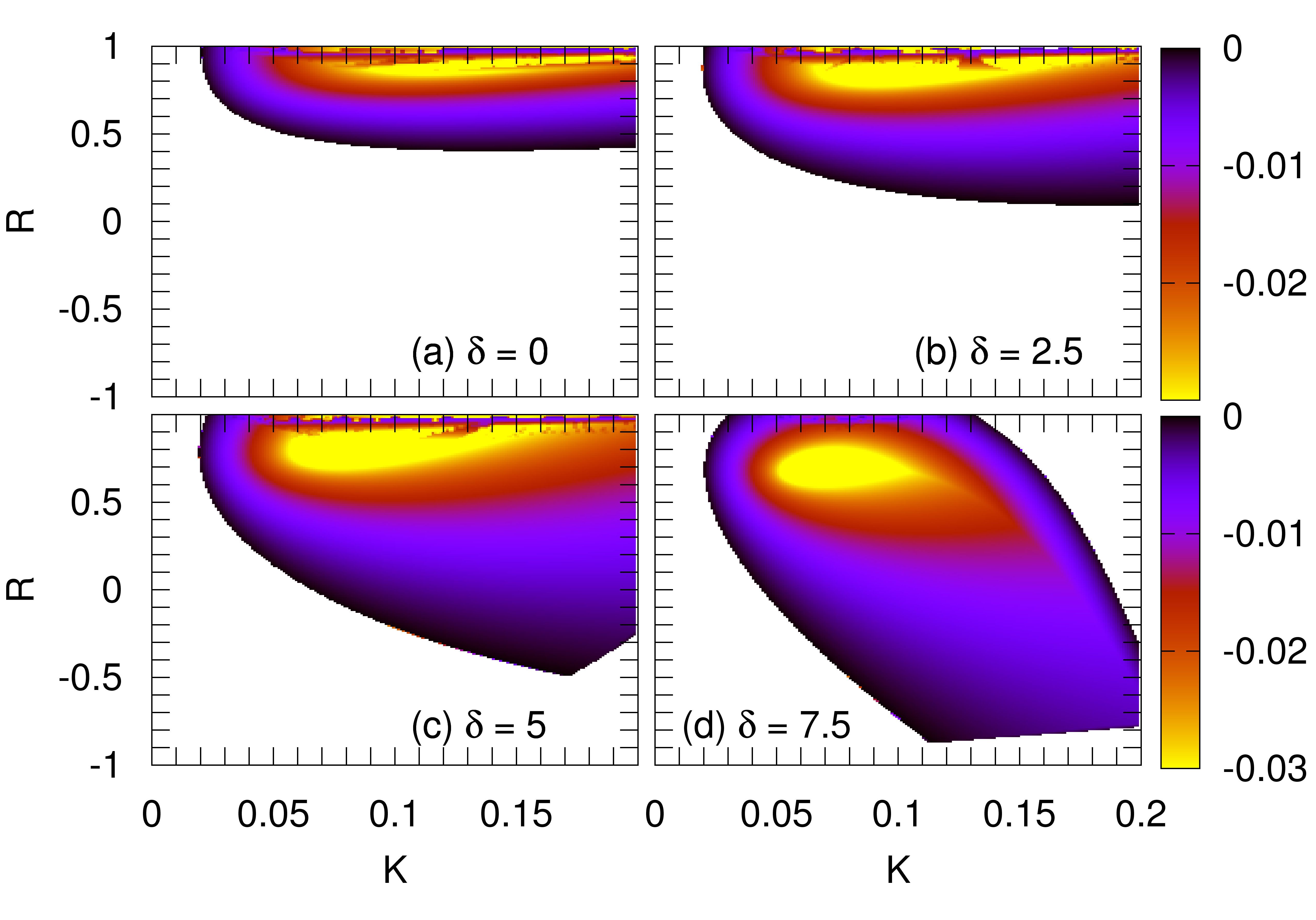}
	\caption{(Color online) Domain of control in the $(K,R)$-plane for different values of $\delta$.  The time delay
and feedback phase are fixed at $\tau=5$ and $\varphi=0$, respectively. The greyscale (color code) denotes the largest
real part $\mathrm{Re}(\Lambda)$ of the eigenvalues $\Lambda$, only negative values are plotted. Panels (a), (b), (c),
and (d) correspond to $\delta=0$, $2.5$, $5$, and $7.5$, respectively.  Other parameters as in
Fig.~\ref{fig:lk_k_tau_domain_phi}.}
	\label{fig:lk_k_r_domain_delta2}
\end{figure}
Previously, it was shown in the generic model of an unstable focus \cite{DAH07} that proper tuning of the latency can
compensate for a bad choice of the time delay $\tau$. Therefore, we depict the domain of control in the $(K,R)$-plane in
\figref{fig:lk_k_r_domain_delta2} for a value of $\tau=5$ and the same values of $\delta$ as in
\figref{fig:lk_k_r_domain_delta}. Note that the domain of control is smaller than in \figref{fig:lk_k_r_domain_delta}
for $\delta=0$. However, the size of the domain of control is greatly enhanced for increasing latency time. The region
of stability is bent towards smaller values of the memory parameter $R$, allowing for control at a smaller value of the
feedback gain for given $R$. The region of optimum stability is located at a large value of $R$ for $\delta=0$. This
region also moves towards lower values of $R$ and grows for increasing $\delta$.

\begin{figure}[ht]
	\includegraphics[width=1.0\columnwidth]{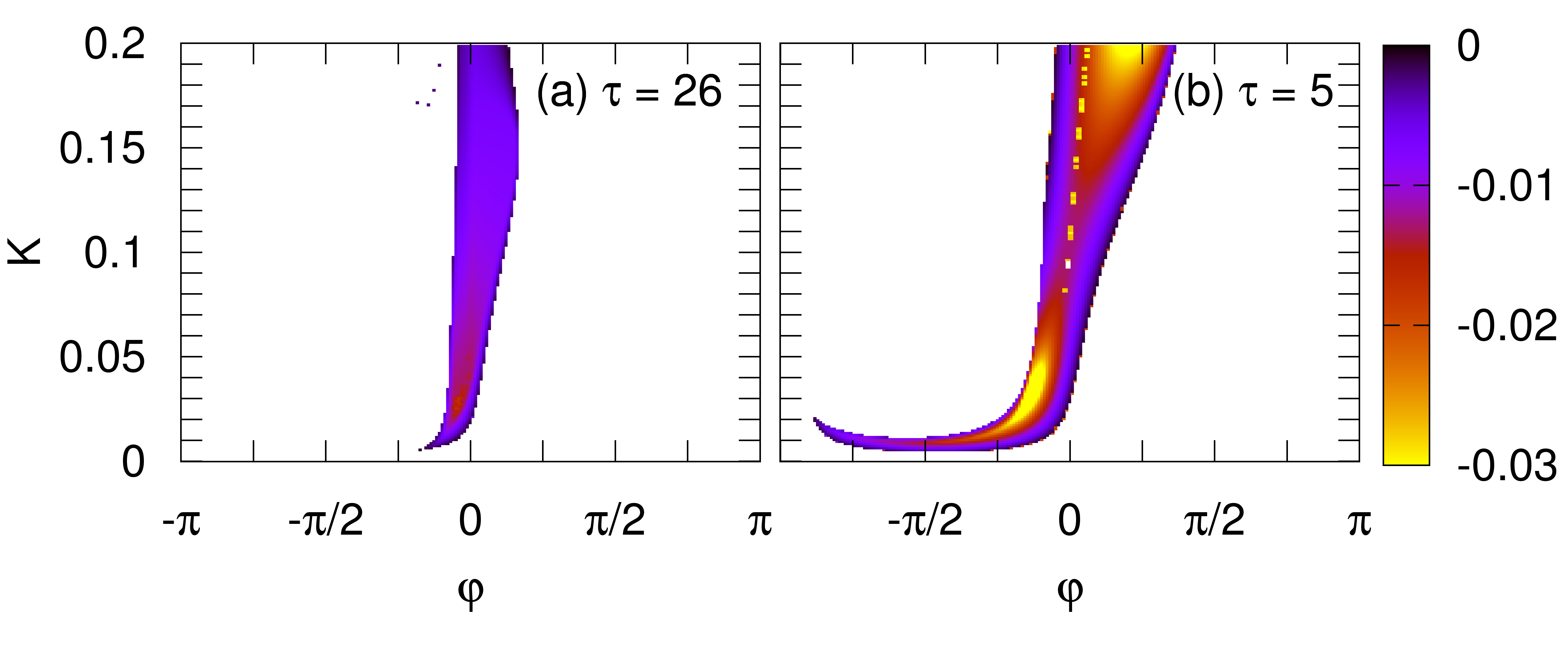}
	\caption{(Color online) Domain of control in the $(K,\varphi)$-plane for different values of $\tau$ and fixed
$\delta=0$. The greyscale (color code) denotes the largest real part $\mathrm{Re}(\Lambda)$ of the eigenvalues
$\Lambda$, only negative values are plotted. Panels (a) and (b) correspond to $\tau=26$ and $5$, respectively. Other
parameters as in Fig.~\ref{fig:lk_k_tau_domain_phi}.}
	\label{fig:lk_k_phi_domain}
\end{figure}
To investigate the dependence of the domain of control on the choice of the phase $\varphi$ further, we consider another
two-dimensional projection of the five-dimensional control-parameter space parameterized by feedback gain $K$ and the
feedback phase $\varphi$. This section is depicted in  \figref{fig:lk_k_phi_domain} for two different values of the time
delay and fixed $\delta=0$. In panel (a), the time delay is chosen as $26$, which is the optimum $\tau$ according to
\eqnref{eqn:tau_opt}. Here, it can be seen that the optimum phase is located at slightly negative values for small
values of the feedback gain up to $K\approx 0.05$. Increasing $K$, the optimum phase changes its sign and is now located
at small positive values of $\varphi$. For the case of $\tau=5$, which is depicted in panel (b), stability is overall
enhanced drastically. The bright (yellow) areas, denoting regions of optimum stability, are located at negative
$\varphi$ for small $K$ up to $K \approx 0.1$. Control is possible even for a small value of $\varphi$ below
$-\pi/2$, if the feedback gain is tuned exactly to the small range of $K \approx 0.01$. For larger feedback gain with
$K>0.1$, the optimum value of $\varphi$ is located at positive values. The region of optimum stability is located at
large values of $K$ around $0.2$. The shape of the control domain in \figref{fig:lk_k_phi_domain} is markedly different
from that in the generic normal form model (see Figs.~6(c) and 7(c) in Ref.~\cite{DAH07}), but appears to be in line
with full device simulations within a travelling wave model \cite{WUE07}.

\section{Bandpass filtering of the feedback signal}
\label{sec:bandpass}
A bandpass filter is widely used in experiments with optical systems, as well as in theoretical treatments of the LK
model \cite{ERZ06, FIS00a}. It is used to suppress unwanted frequencies in the feedback loop but can also
fundamentally influence the dynamics of the laser system \cite{FIS04}. Therefore it is useful to include a corresponding
term into the theory. Experimentally, the filter is realized by a Fabry-Perot interferometer, which can
approximatively be modelled by a Lorentzian. Therefore, the bandpass filter is introduced by using the transfer function $T(\omega)$ of
a Lorentzian in Fourier space. The transfer function acting on the electric field $E$ can be written as
\begin{equation}
 T(\omega) = \frac{1}{1+\imath\frac{\omega-\omega_{0}}{\gamma}},
\end{equation}
where $\omega_0$ denotes the peak of the transfer function and $\gamma$ the full width at half maximum. In Fourier
space, the bandpass filter alters the complex electric field $E(t)$ as follows
\begin{eqnarray}\label{eqn:lk_filtermittransfer}
 \bar{E}(\omega) & = & T(\omega) E(\omega) \nonumber \\
 & = & \frac{1}{1+\imath\frac{\omega-\omega_{0}}{\gamma}} E(\omega)
\end{eqnarray}
with the filtered electric field  $\bar{E}$. Transforming back from Fourier space yields a differential equation for
$\bar{E}(t)$:
\begin{eqnarray} \label{eqn:lk_filter_dgl_fuer_E}
 \frac{d \bar{E}(t)}{dt} &=& \left( \imath \omega_{0} - \gamma \right) \bar{E}(t) + \gamma E(t),
\end{eqnarray}
which is added to the model equations \eqref{eqn:lk_tronciu},
\begin{subequations}
\label{eqn:lk_tronciu_bandpass}
\begin{eqnarray}
  \frac{d E}{d t} & = & \frac{T}{2} \left( 1+ \imath \alpha \right) n E - E_{b}(t) ,\\
  \frac{d n}{d t} & = & I -n- \left[1+n\right] k(n) \left| E \right|^{2}, \\
  \frac{d \bar{E}}{d t} &=& \left( \imath \omega_{0} - \gamma \right) \bar{E}(t) + \gamma E(t),
\end{eqnarray}
\end{subequations}
while the filtered field $\bar{E}(t)$ is used instead of $E(t)$ in the feedback term \eqref{eqn:lk_fp_feedback}:
\begin{eqnarray}
 \lefteqn{ E_{b}(t)} \label{eqn:lk_fp_feedback_bandpass}\\
 & = & K e^{-i\varphi} \sum_{m=0}^{\infty} R^{m} \left[ \bar{E}(t-\delta-m \tau) - \bar{E}(t-\delta-(m+1)\tau) \right]
\nonumber
\end{eqnarray}
Similar to the unfiltered system, the complex filtered variable $\bar{E}(t)$ can be split into real and imaginary parts
denoted by $\bar{x}(t)$ and $\bar{y}(t)$, respectively.
This leads to additional differential equations for the filtered variables $\bar{x}(t)$ and $\bar{y}(t)$:
\begin{subequations}
\label{eqn:lk_xyquer}
\begin{eqnarray}
\frac{d \bar{x}(t)}{dt} & = & \gamma \left( x(t) - \bar{x}(t) \right) - \omega_{0} \bar{y}(t) ,\\
\frac{d \bar{y}(t)}{dt} & = & \gamma \left( y(t) - \bar{y}(t) \right) + \omega_{0}\bar{x}(t),
\end{eqnarray}
\end{subequations}
where we rescaled the parameters $\gamma$ and $\omega_0$ by $2/T$. The time has also been rescaled as in
Eq.~(\ref{eqn:lk_linear}).

For the sake of simplicity, we restrict ourselves to the case of $\varphi=0$ and $\delta=0$ in the linear stability
analysis, which corresponds to the case of zero phase and no additional latency. 
Using Eqs.~\ref{eqn:lk_xyquer} for the filtered variables $\bar{x}$ and $\bar{y}$ and the linearization from
\eqnref{eqn:lk_linear}, the characteristic equation of the controlled system with bandpass filter can be obtained:
\begin{widetext}
\begin{eqnarray}
0 &=& 4 \alpha \gamma \omega_{0} \Omega_{0}^{2} K \frac{1-e^{-\Lambda \tau}}{1- R e^{-\Lambda \tau}} +\Lambda
\omega_{0}^{2} \left( 2 \Gamma \Lambda + \Lambda^{2} + 4 \Omega_{0}^{2}\right) + \left[\gamma K \frac{1-e^{-\Lambda
\tau}}{1- R e^{-\Lambda \tau}} + \Lambda \left(\gamma+\Lambda\right)\right]  \nonumber \\
& & \times \left\{ - \left(2\Gamma +\Lambda\right) \left[\gamma K \frac{1-e^{-\Lambda \tau}}{1- R e^{-\Lambda \tau}} +
\Lambda \left(\gamma+\Lambda\right)\right] - 4 \Omega_{0}^{2}\left(\gamma+\Lambda\right) \right\} .
\label{eqn:bandpass_ev}
\end{eqnarray}
\end{widetext}
The eigenvalues $\Lambda$, which are the roots of the characteristic equation, are calculated numerically.
Details on the numerical procedure are given in the beginning of
Sec.~\ref{sec:results}.
The parameter space is five-dimensional and parameterized by $K$, $\tau$, $R$, $\gamma$, and $\omega_{0}$. Results are shown
in projections onto the ($K$,$\tau$) parameter plane in the following.

\begin{figure}[ht]
	\includegraphics[width=1.0\columnwidth]{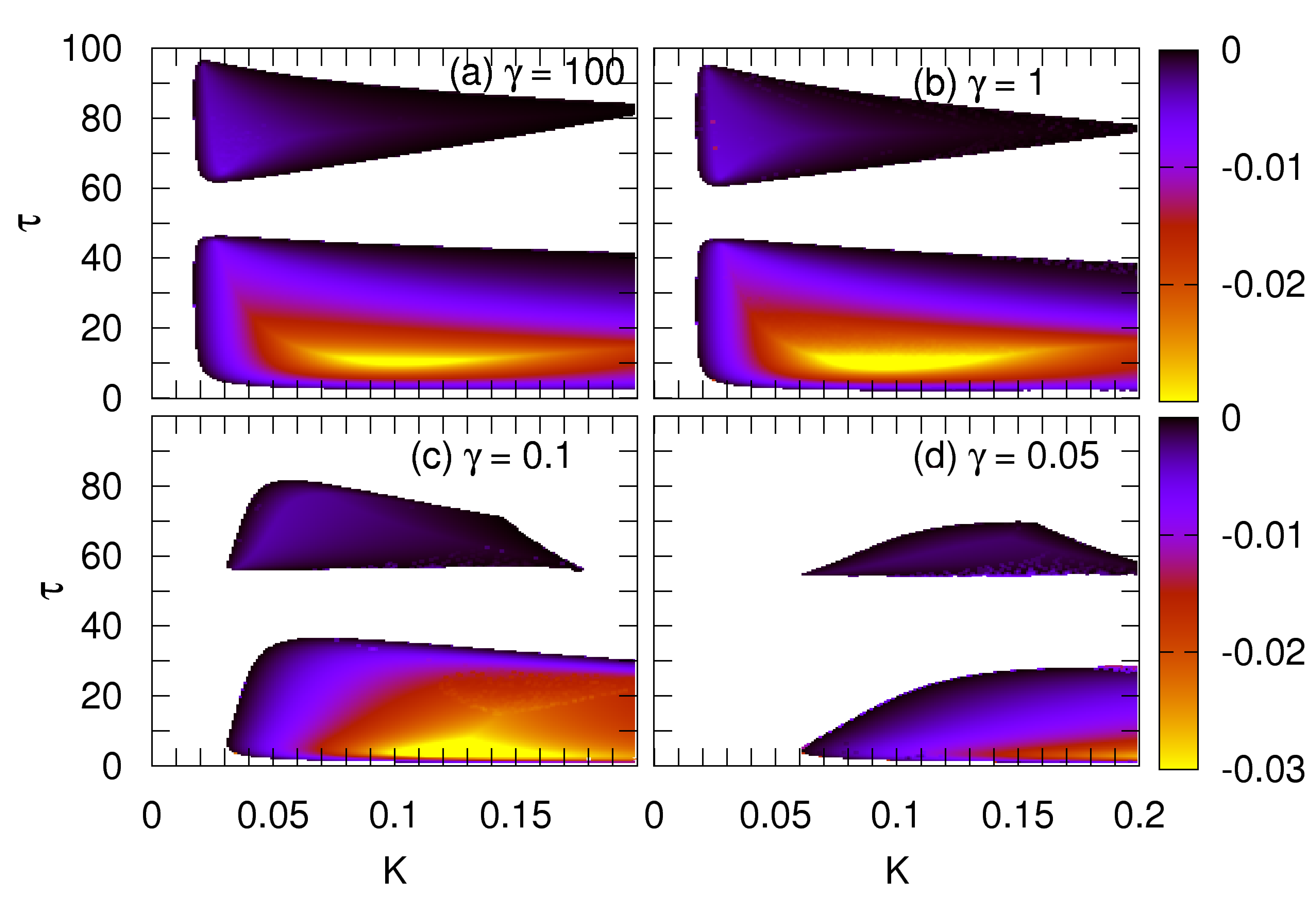}
	\caption{(Color online) Domain of control in the $(K,\tau)$-plane according to Eq.~\eqref{eqn:bandpass_ev} for
different values of $\gamma=100$, $1$, $0.1$, and $0.05$ in panels (a), (b), (c), and (d), respectively. $\omega_{0}=0$
is fixed. The greyscale (color code) denotes the largest real part $\mathrm{Re}(\Lambda)$ of the eigenvalues $\Lambda$,
only negative values are plotted. Other parameters as in Fig.~\ref{fig:lk_k_tau_domain_phi}.}
\label{fig:k_tau_domain_lk_ev_gamma}
\end{figure}
In \figref{fig:k_tau_domain_lk_ev_gamma}, the domain of control is shown in the $(K,\tau)$-plane for fixed
$\omega_{0}=0$ and different values of the filter width: $\gamma=100$, $1$, $0.1$, and $0.05$. The choice of $\omega_0$
corresponds to the case of the low-pass filter because the peak of the transfer function is shifted to zero. The
greyscale (color code) corresponds to the largest real part of the eigenvalues. Only negative values are displayed. For
large values of $\gamma$, the picture is almost identical to the case of the unfiltered system shown in panel (a) of
\figref{fig:lk_k_tau_domain_phi}. This can easily be understood because, for large cut-off frequency $\gamma$, the
electric field passes the filter almost unchanged. Decreasing the filter width to $\gamma=1$ has only little effect on
the shape of the domain of control, but the region of best stability, denoted by bright (yellow) color, becomes larger.
Decreasing $\gamma$ further, the domain of control shrinks and is bent down slightly towards smaller values of $\tau$.
This suggests that the optimum $\tau$ changes, which will be investigated later in the $(K,R)$-plane. Further increase
of $\gamma$ leads to smaller domains because higher frequency components are cut off and thus, more information of the
system is lost due to the filter.

\begin{figure}[ht]
	\includegraphics[width=1.0\columnwidth]{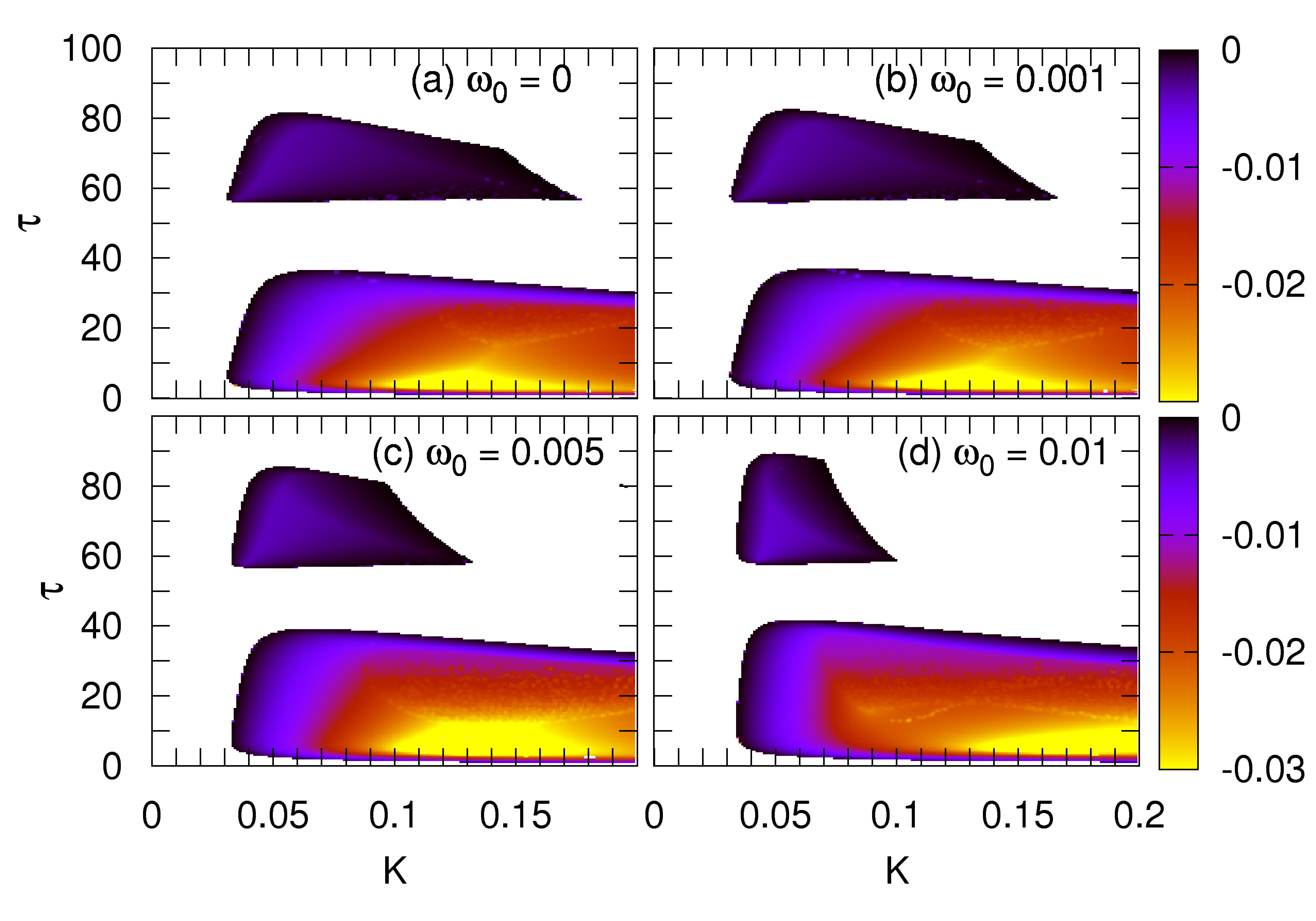}
	\caption{(Color online) Domain of control in the $(K,\tau)$-plane according to Eq.~\eqref{eqn:bandpass_ev} for different values of $\omega_{0}$ with fixed $\gamma=0.1$. The greyscale (color code) denotes the largest real part $\mathrm{Re}(\Lambda)$ of the eigenvalues $\Lambda$, only negative values are plotted. Parameters: $\Gamma=-0.01$, $\Omega_{0}=0.06$, $\alpha=5$, and $R=0.7$. Panels (a), (b), (c), and (d) correspond to $\omega_{0}=0$, $0.001$, $0.005$, and $0.01$, respectively.}
	\label{fig:k_tau_domain_lk_ev_omega0}
\end{figure}
\figureref{fig:k_tau_domain_lk_ev_omega0} depicts the domain of control in the $(K,\tau)$-plane for different values of
the filter center frequency $\omega_{0}=0$, $0.001$, $0.005$, and $0.01$ in panels (a), (b), (c), and (d), respectively.
The value of $\gamma$ is fixed at $\gamma = 0.1$. The increase of $\omega_{0}$ has only little effect on the first
tongue of stability. However, the second tongue is cut off from the upper right with larger $\omega_{0}$, leading to a
smaller range of possible values for the feedback gain $K$. This tongue gets also slightly thicker in $\tau$ direction
for increasing $\omega_{0}$.

\section{Delay-induced multistability}
\label{sec:lk_multistability}
In all preceding considerations, the effect of time-delayed feedback control on the stability of the lasing fixed point
was investigated only locally, since the system was linearized around the fixed point. Results from dynamical
simulations of the full nonlinear system, given by 
Eqs.~(\ref{eqn:lk_tronciu}), with feedback according to
\eqnref{eqn:lk_fp_feedback}, are reported in this Section.

\begin{figure}[ht]
 \begin{center}
	\includegraphics[width=1.0\columnwidth]{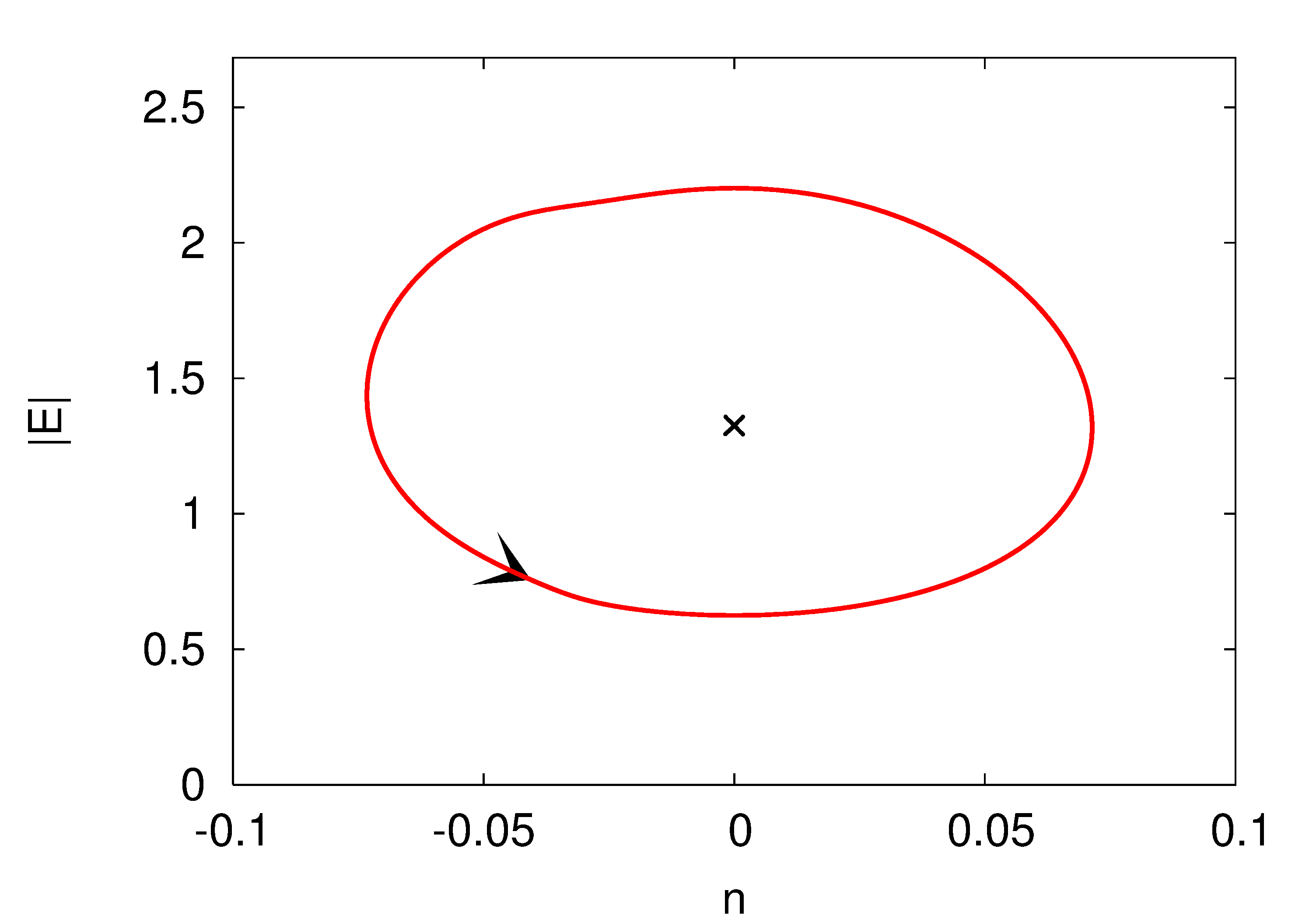}
 \end{center}
 \caption{Absolute value of the electric field vs reduced carrier density: Limit cycle of the undamped relaxation
oscillations in the modified LK model according to \eqnref{eqn:lk_tronciu}
and \eqnref{eqn:lk_kvonn} without feedback. Parameters: $T=500$, $\alpha=5$, $I=1.8$, $A=1$, $W=0.02$, and $n_0 = -0.034$. The arrow indicates the direction of the trajectory. The cross denotes the lasing (cw) fixed point.}
 \label{fig:lk_limitcycle}
\end{figure}
Besides the fixed point associated with cw emission, the model without control, described by Eqs.~(\ref{eqn:lk_tronciu}), has a limit cycle in the $(|E|,n)$-plane, which corresponds to undamped relaxation oscillations (or intensity pulsations). The amplitude of this limit cycle, which is born by a Hopf bifurcation at $\Gamma=0$, depends on the choice of the parameters. For the same parameters as in Sec.~\ref{sec:linear}, the limit cycle is depicted in \figref{fig:lk_limitcycle}. The unstable fixed point is located at $(n=0, |E|=1.342)$ for this choice of parameters.

In an experiment, the case may be of interest where the feedback is activated when the laser is operating at this limit
cycle of relaxation oscillations. Therefore, for numerical simulations we choose initial conditions on this limit cycle:
$E(t=0)=-0.731767-i\;0.016891$, which corresponds to $x(0)=-0.0927256$ and $y(0)=-0.00075538$. The initial values of
the reduced carrier density is chosen as $n(0)=-0.037433$.

\begin{figure}[ht]
	\includegraphics[width=1.0\columnwidth]{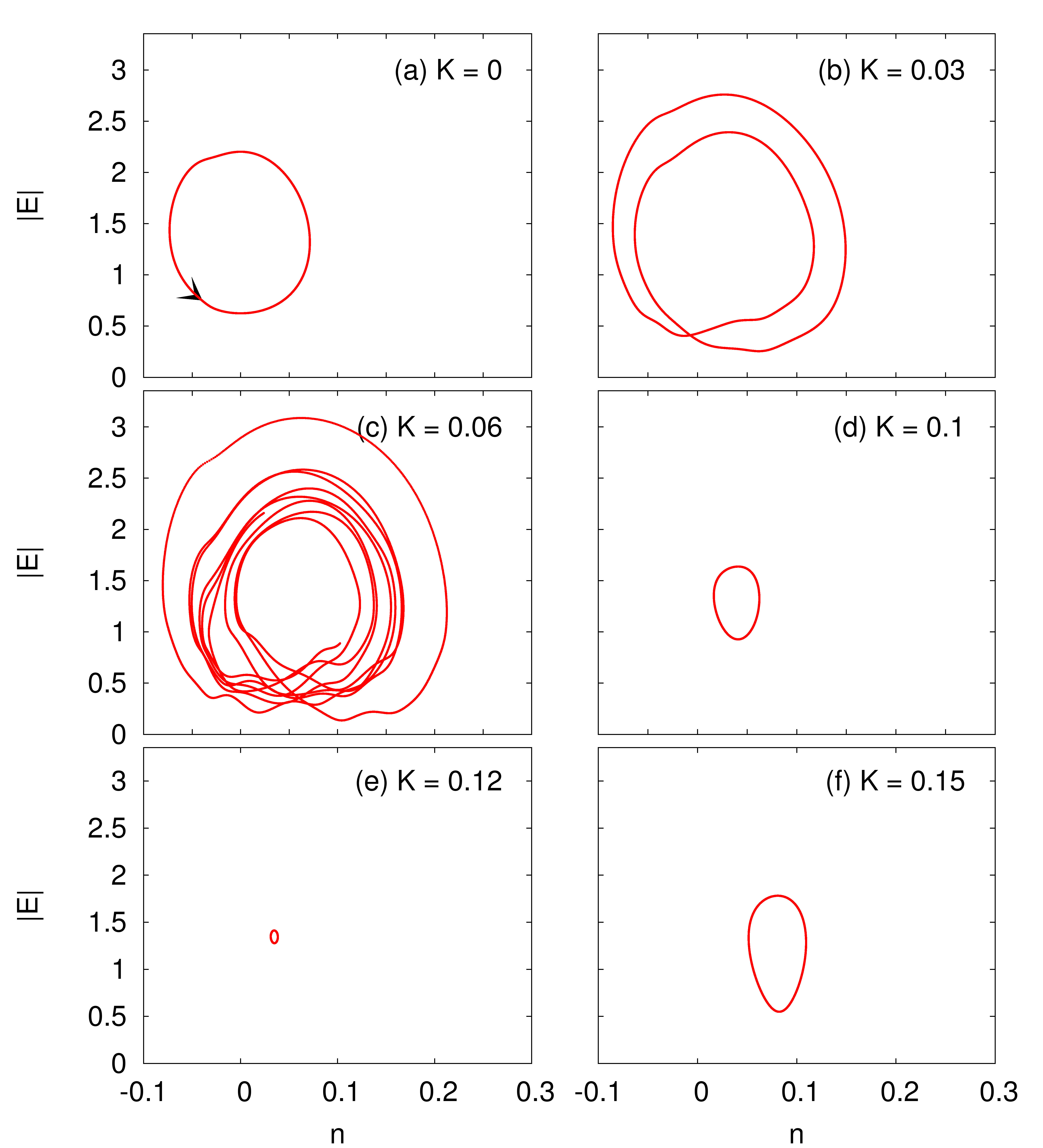}
 \caption{Dynamics with feedback for different values of the feedback gain $K$. Initial conditions are chosen on the
limit cycle of relaxation oscillations (Fig.~\ref{fig:lk_limitcycle}). The time delay and memory parameter are fixed at
$\tau=26$ and $R=0.7$, respectively. $\varphi=0$, $\delta=0$. Panels (a), (b), (c), (d), (e), and (f) correspond to
$K=0$, $0.03$, $0.06$, $0.1$, $0.12$, and $0.15$, respectively. The rotational direction of the trajectories is the same
as in (a) for all other panels. Other parameters as in Fig.~\ref{fig:lk_limitcycle}. Note that the trajectories are
shown only for a limited span of time.}
 \label{fig:lk_traj_verschk}
\end{figure}

In the presence of a feedback term, the system equations become a set of delay differential equations. Therefore, the initial conditions must be specified for the time interval $[-\tau,0)$. For this interval, we choose the values of the limit cycle shown in Fig.~\ref{fig:lk_limitcycle}. Depending on the feedback gain $K$ and the time delay $\tau$, the system exhibits diverse scenarios under the influence of feedback according to \eqnref{eqn:lk_fp_feedback}. Some trajectories of the system in the $(|E|,n)$-plane are displayed in \figref{fig:lk_traj_verschk} for fixed time delay $\tau=26$ and different values of the feedback gain $K$. 
The memory parameter is fixed to $R=0.7$. A variety of dynamic scenarios ranging from limit cycles to chaotic attractors
can be observed depending on the choice of $K$. Panel (a), in the absence of control ($K=0$), shows the limit cycle of
undamped relaxation oscillations as \figref{fig:lk_limitcycle}. Increasing the feedback gain to $K=0.03$ (b), a limit
cycle of period $2$ evolves. For a value of $K=0.06$ (c), chaotic behavior is observed. At $K=0.1$ (d) a small limit
cycle appears, but this limit cycle does not oscillate around the uncontrolled fixed point, which is located at $n=0$.
Upon further increase the limit cycle shrinks (e), and finally disappears in an inverse Hopf bifurcation from a
different fixed point. A different limit cycle is observed for $K=0.15$ (f). For other parameters, a variety of complex
scenarios is found. This is consistent with recent findings by Tronciu \textit{et al.} \cite{TRO06}, who demonstrated
rich multistable dynamic scenarios of feedback-induced stationary external cavity modes (rotating waves $E(t)=E_S e^{i
\omega_S t}$ and further Hopf bifurcations of those) if a certain critical feedback strength $K_c$ is exceeded.
Note that the lasing fixed point (solitary laser mode) at $n=0$ is stable for all values of
$K>0$ in 
Fig.~\ref{fig:lk_traj_verschk}, thus there exists multistability between cw laser emission and (periodic or chaotic) intensity pulsations, and it depends upon the initial conditions which attractor is asymptotically reached. If the initial condition is chosen in the vicinity of the fixed point, where the linearization is valid, time-delayed feedback control still works to stabilize cw emission. This can be achieved, for instance, by first operating the laser below the Hopf bifurcation and then gradually increasing the pump current, and hence $\Gamma$, as will be shown in the next Section. 
Note that delay-induced multistability is common in many systems \cite{HIZ07}.

\section{Turn-on dynamics}
\label{sec:turnon}
The results of the preceding Section might suggest that stabilization of the lasing fixed point (cw emission) in the full nonlinear system is difficult if not impossible for the majority of choices for the control parameters $K$, $\tau$, $R$, $\delta$, $\phi$, $\gamma$, and $\omega_{0}$. Driving the system into the lasing fixed point is only possible for initial conditions in a vicinity of the lasing fixed point. An experimental realization of such initial conditions will most probably be complicated. Choosing other initial conditions, for example on the limit cycle of relaxation oscillations, as shown in the previous Section, leads to unwanted multistability in the presence of control.

We have therefore simulated a turn-on scenario which is experimentally both realistic and feasible. As a result of the slowly increasing instability the feedback is able to keep the dynamics in the very vicinity of the lasing fixed point.
Even with a small amount of noise this behavior is expected to persist, 
since the basin of attraction of the fixed point is finite.

\begin{figure}[ht]
	\includegraphics[width=1.0\columnwidth]{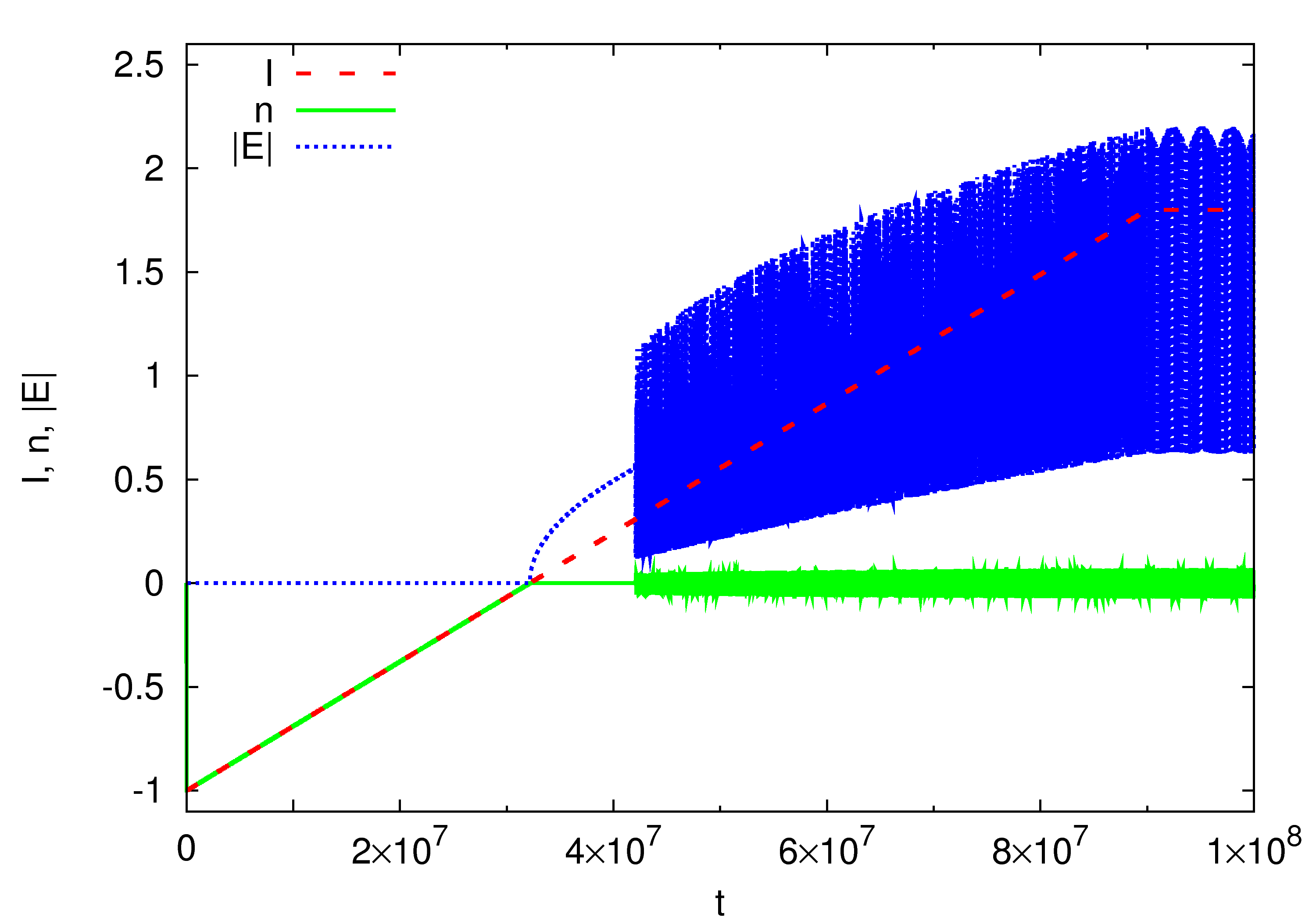}
	\caption{(Color online) Dynamics of the system during a turn-on process without feedback ($K=0$). The dashed
(red), solid (green), and dotted (blue) curves correspond to the injection current $I$, the carrier inversion $n$, and
the absolute value of the electric field $|E|$, respectively, vs rescaled dimensionless time. Other parameters as in
Fig.~\ref{fig:lk_k_tau_domain_phi}.}
	\label{fig:lk_turnon_K0}
\end{figure}
The turn-on scenario is depicted in Fig.~\ref{fig:lk_turnon_K0} for the case of no control. The laser is first operated
below the laser threshold with the injection current chosen as $I=-1$. With initial conditions chosen arbitrarily as
$x=y=10^{-6}$ and $n=0$ the laser relaxes into the non-lasing fixed point. The injection current is then linearly
increased until the value of $I=1.8$ is reached, which was considered in the previous Sections of this work. Passing the
laser threshold at $I=0$, non-lasing and lasing fixed points change stability, thus the system swaps to the lasing fixed
point with $n=0$ and nonzero $E$. Increasing the current further beyond the value of $I=0.3$, the lasing fixed point
becomes again unstable due to the internal dispersive feedback (see Eq.~(\ref{eqn:lk_eigenvalues_uncontrolled})). 
Reaching $I=1.8$, the system resides in the limit cycle shown in Fig.~\ref{fig:lk_limitcycle}. 
The full turn-on process from $I=-1$ to $1.8$ takes $9 \times 10^{7}$ 
in the rescaled time units according to Eq.~(\ref{eq:scale}) and used in these simulations. Assuming a photon lifetime
of the order of 
$\tau_{p} \approx 10^{-11}$s, this would relate to a physical time of 
$s_{\textnormal{turn-on}}\approx 1.8$ms for the turn-on ramp using $s=
(2\tau_{c}/T)t= 2 \tau_{p} t$, where $t$ is the rescaled dimensionless time and $s$ is the time with physical units.
Note that $\tau_{c}=5 \times 10^{-9}$s for the chosen values of $T$ and
$\tau_{p}$. 

\begin{figure}[ht]
	\includegraphics[width=1.0\columnwidth]{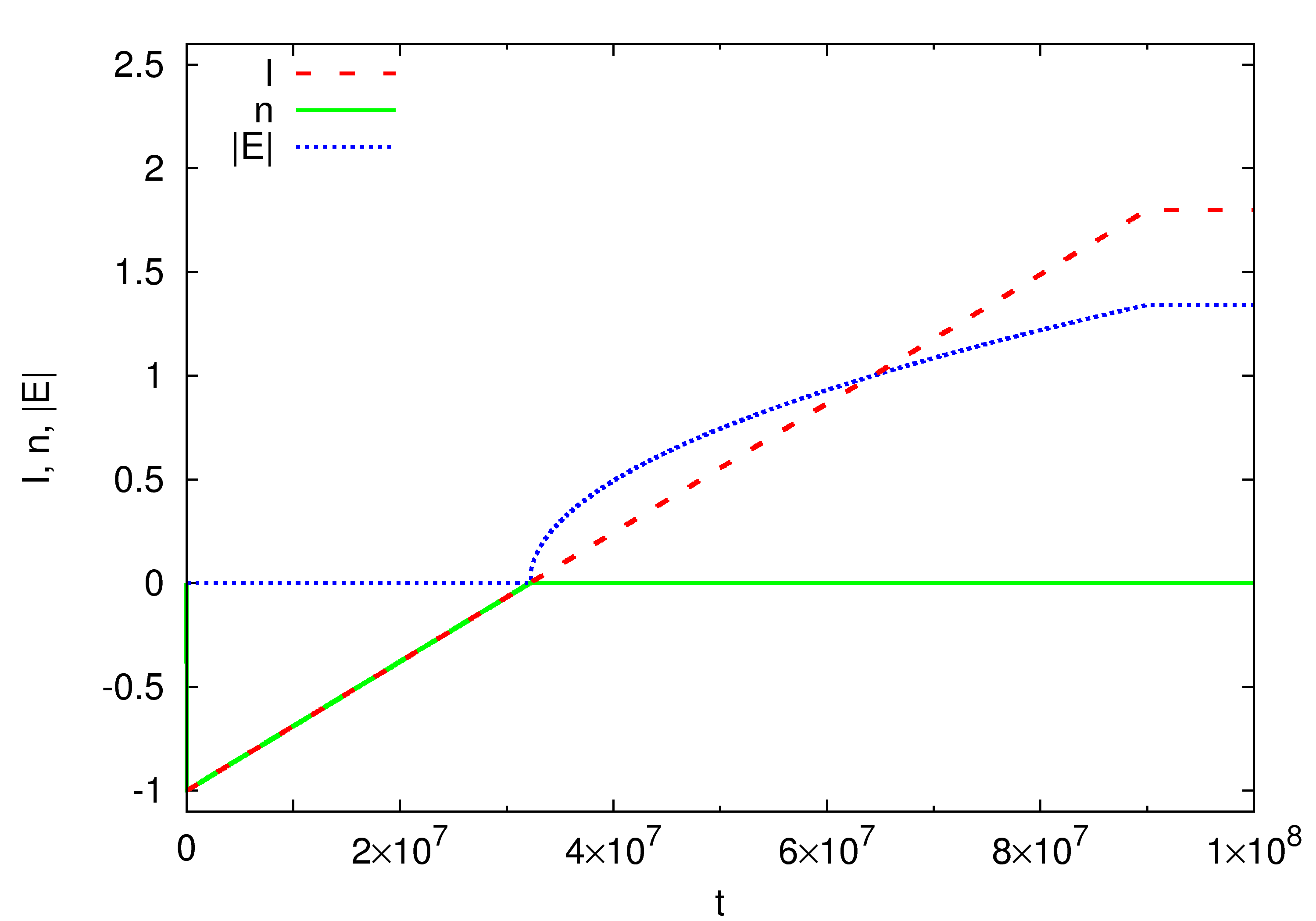}
	\caption{(Color online) Dynamics of the system during a turn-on process with feedback ($K=0.1$, $\tau=26$). The
dashed (red), solid (green), and dotted (blue) curves correspond to the injection current $I$, the carrier inversion
$n$, and the absolute value of the electric field $|E|$, respectively. Other parameters: $R=0.7$, $\varphi=0$, and
$\delta=0$.}
	\label{fig:lk_turnon_K0.1}
\end{figure}
We now consider the system in the presence of control. For experimental feasibility the feedback is active during the
full turn-on process. We consider an exemplary case with $\tau=26$, $K=0.1$, and $R=0.7$. Phase and latency are chosen
as zero, the bandpass filter is turned off. The dynamics during the turn-on process is shown in
Fig.~\ref{fig:lk_turnon_K0.1}. It can be seen that the system remains in the vicinity of the lasing fixed point even
beyond the point $I=0.3$. This is achieved since the feedback is already active as the system passes this onset of
instability. As the system is thereby always kept near the fixed point the results of the linear stability analysis of
Section~\ref{sec:linear} can be applied here. Another evidence for the success of the control is the fact that after the
turn-on process the control force vanishes. In the simulations of Section~\ref{sec:lk_multistability}, where only
delay-induced orbits were created, the control force did not vanish.

\begin{figure}[ht]
	\includegraphics[width=1.0\columnwidth]{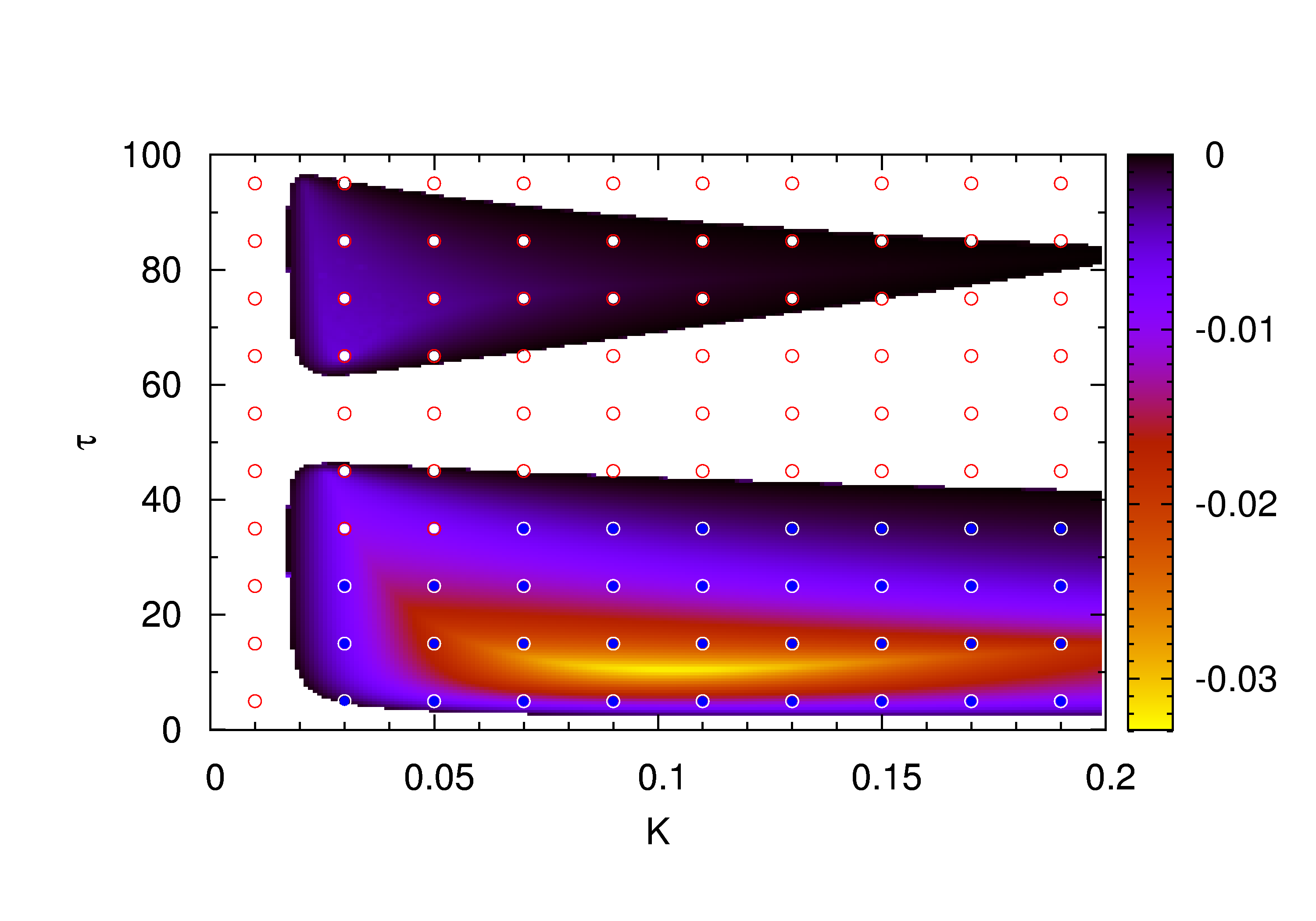}
	\caption{(Color online) Domain of control according to Eq.~\eqref{eqn:EIGENVALUES_LK} in the $(K,\tau)$-plane compared to results from turn-on simulations. The greyscale (color code) denotes the largest real part $\mathrm{Re}(\Lambda)$ of the eigenvalues $\Lambda$, only negative values are plotted. Circles show results from simulations. Full (blue) circles correspond to successful control of the fixed point, empty (red) circles denote no control. Parameters: $\alpha=5$, $R=0.7$, $\varphi=0$, and $\delta=0$.}
	\label{fig:lk_k_tau_domain_sim}
\end{figure}

In Fig.~\ref{fig:lk_k_tau_domain_sim}, the analytic results from Section~\ref{sec:linear} are plotted together with the results from turn-on simulations as shown before.
In the lower part of the picture for relatively small values of $\tau$, the results from the simulations match the results from the linear stability analysis. However, in the upper part of the lower tongue of control and in the whole upper tongue of control, the simulations are not successful. The large value of the time delay makes control less effective at the onset of instability of the fixed point (at $I=0.3$). This drives the trajectory slightly away from the fixed point, which can afterwards not be corrected by the control. Choosing a slower ramp for the turn-on process retains control for larger values of $\tau$. We have performed exemplary simulations with a turn-on time ten times longer than before (now $9 \times 10^{8}$ in rescaled dimensionless units) using $K=0.03$ for
a different value of the time delay. For $\tau=35$, control was then successful.

\section{Conclusion}
\label{sec:conclusion}
In conclusion, we have shown that time-delayed feedback provides a valuable tool for the suppression of unwanted intensity pulsations in a semiconductor laser, which can conveniently be realized by a Fabry-Perot resonator. We stress that the efficiency of this method has already been demonstrated experimentally \cite{SCH06a}. We have discussed the effects of the control scheme in the framework of a modified Lang-Kobayashi model. Our results were obtained by a linear stability analysis of the fixed point of the delayed system. As a modification of the original controller design, we have taken into account an additional control loop latency as well as a variable phase-dependent coupling. We have shown how these extensions affect the domain of control in various projections of the parameter space. Futhermore, we have investigated the effects of a bandpass filter added in the feedback. In addition, we have presented simulations of the full nonlinear system, which may exhibit multistability, so that simple feedback control fails. In this case, control may be achieved only by slowly increasing the injection current until the desired value is reached. This corresponds experimentally to turn-on of the laser with a slow injection current ramp.

\section{Acknowledgements}
This work was supported by Deutsche Forschungsgemeinschaft in the framework of Sfb 555. We thank S. Schikora, H.-J.
W{\"u}nsche, and V. Z. Tronciu for valuable discussions.


\end{document}